\documentstyle[aps,floats,prb]{revtex}
\edef\psfigRestoreAt{\catcode`@=\number\catcode`@\relax}
\catcode`\@=11\relax
\newwrite\@unused
\def\typeout#1{{\let\protect\string\immediate\write\@unused{#1}}}
\def\figurepath{./}

\def\@nnil{\@nil}
\def\@empty{}
\def\@psdonoop#1\@@#2#3{}
\def\@psdo#1:=#2\do#3{\edef\@psdotmp{#2}\ifx\@psdotmp\@empty \else
    \expandafter\@psdoloop#2,\@nil,\@nil\@@#1{#3}\fi}
\def\@psdoloop#1,#2,#3\@@#4#5{\def#4{#1}\ifx #4\@nnil \else
       #5\def#4{#2}\ifx #4\@nnil \else#5\@ipsdoloop #3\@@#4{#5}\fi\fi}
\def\@ipsdoloop#1,#2\@@#3#4{\def#3{#1}\ifx #3\@nnil 
       \let\@nextwhile=\@psdonoop \else
      #4\relax\let\@nextwhile=\@ipsdoloop\fi\@nextwhile#2\@@#3{#4}}
\def\@tpsdo#1:=#2\do#3{\xdef\@psdotmp{#2}\ifx\@psdotmp\@empty \else
    \@tpsdoloop#2\@nil\@nil\@@#1{#3}\fi}
\def\@tpsdoloop#1#2\@@#3#4{\def#3{#1}\ifx #3\@nnil 
       \let\@nextwhile=\@psdonoop \else
      #4\relax\let\@nextwhile=\@tpsdoloop\fi\@nextwhile#2\@@#3{#4}}
\newread\ps@stream
\newif\ifnot@eof       
\newif\if@noisy        
\newif\if@atend        
\newif\if@psfile       
{\catcode`\%=12\global\gdef\epsf@start{
\def\epsf@PS{PS}
\def\epsf@getbb#1{%
\openin\ps@stream=#1
\ifeof\ps@stream\typeout{Error, File #1 not found}\else
   {\not@eoftrue \chardef\other=12
    \def\do##1{\catcode`##1=\other}\dospecials \catcode`\ =10
    \loop
       \if@psfile
	  \read\ps@stream to \epsf@fileline
       \else{
	  \obeyspaces
          \read\ps@stream to \epsf@tmp\global\let\epsf@fileline\epsf@tmp}
       \fi
       \ifeof\ps@stream\not@eoffalse\else
       \if@psfile\else
       \expandafter\epsf@test\epsf@fileline:. \\%
       \fi
          \expandafter\epsf@aux\epsf@fileline:. \\%
       \fi
   \ifnot@eof\repeat
   }\closein\ps@stream\fi}%
\long\def\epsf@test#1#2#3:#4\\{\def\epsf@testit{#1#2}
			\ifx\epsf@testit\epsf@start\else
\typeout{Warning! File does not start with `\epsf@start'.  It may not be a PostScript file.}
			\fi
			\@psfiletrue} 
{\catcode`\%=12\global\let\epsf@percent=
\long\def\epsf@aux#1#2:#3\\{\ifx#1\epsf@percent
   \def\epsf@testit{#2}\ifx\epsf@testit\epsf@bblit
	\@atendfalse
        \epsf@atend #3 . \\%
	\if@atend	
	   \if@verbose{
		\typeout{psfig: found `(atend)'; continuing search}
	   }\fi
        \else
        \epsf@grab #3 . . . \\%
        \not@eoffalse
        \global\no@bbfalse
        \fi
   \fi\fi}%
\def\epsf@grab #1 #2 #3 #4 #5\\{%
   \global\def\epsf@llx{#1}\ifx\epsf@llx\empty
      \epsf@grab #2 #3 #4 #5 .\\\else
   \global\def\epsf@lly{#2}%
   \global\def\epsf@urx{#3}\global\def\epsf@ury{#4}\fi}%
\def\epsf@atendlit{(atend)} 
\def\epsf@atend #1 #2 #3\\{%
   \def\epsf@tmp{#1}\ifx\epsf@tmp\empty
      \epsf@atend #2 #3 .\\\else
   \ifx\epsf@tmp\epsf@atendlit\@atendtrue\fi\fi}

\def\psdraft{
	\def\@psdraft{0}
}
\def\psfull{
	\def\@psdraft{100}
}

\psfull

\newif\if@draftbox
\def\psnodraftbox{
	\@draftboxfalse
}
\@draftboxtrue

\newif\if@prologfile
\newif\if@postlogfile
\def\pssilent{
	\@noisyfalse
}
\def\psnoisy{
	\@noisytrue
}
\pssilent 
\newif\if@bbllx
\newif\if@bblly
\newif\if@bburx
\newif\if@bbury
\newif\if@height
\newif\if@width
\newif\if@rheight
\newif\if@rwidth
\newif\if@clip
\newif\if@verbose
\def\@p@@sclip#1{\@cliptrue}

\def\@p@@sfile#1{\def\@p@sfile{null}%
	        \openin1=#1
		\ifeof1\closein1%
		       \openin1=\figurepath#1
			\ifeof1\typeout{Error, File #1 not found}
			\else\closein1
			    \edef\@p@sfile{\figurepath#1}%
                        \fi%
		 \else\closein1%
		       \def\@p@sfile{#1}%
		 \fi}
\def\@p@@sfigure#1{\def\@p@sfile{null}%
	        \openin1=#1
		\ifeof1\closein1%
		       \openin1=\figurepath#1
			\ifeof1\typeout{Error, File #1 not found}
			\else\closein1
			    \def\@p@sfile{\figurepath#1}%
                        \fi%
		 \else\closein1%
		       \def\@p@sfile{#1}%
		 \fi}

\def\@p@@sbbllx#1{
		\@bbllxtrue
		\dimen100=#1
		\edef\@p@sbbllx{\number\dimen100}
}
\def\@p@@sbblly#1{
		\@bbllytrue
		\dimen100=#1
		\edef\@p@sbblly{\number\dimen100}
}
\def\@p@@sbburx#1{
		\@bburxtrue
		\dimen100=#1
		\edef\@p@sbburx{\number\dimen100}
}
\def\@p@@sbbury#1{
		\@bburytrue
		\dimen100=#1
		\edef\@p@sbbury{\number\dimen100}
}
\def\@p@@sheight#1{
		\@heighttrue
		\dimen100=#1
   		\edef\@p@sheight{\number\dimen100}
}
\def\@p@@swidth#1{
		\@widthtrue
		\dimen100=#1
		\edef\@p@swidth{\number\dimen100}
}
\def\@p@@srheight#1{
		\@rheighttrue
		\dimen100=#1
		\edef\@p@srheight{\number\dimen100}
}
\def\@p@@srwidth#1{
		\@rwidthtrue
		\dimen100=#1
		\edef\@p@srwidth{\number\dimen100}
}
\def\@p@@ssilent#1{ 
		\@verbosefalse
}
\def\@p@@sprolog#1{\@prologfiletrue\def\@prologfileval{#1}}
\def\@p@@spostlog#1{\@postlogfiletrue\def\@postlogfileval{#1}}
\def\@cs@name#1{\csname #1\endcsname}
\def\@setparms#1=#2,{\@cs@name{@p@@s#1}{#2}}
\def\ps@init@parms{
		\@bbllxfalse \@bbllyfalse
		\@bburxfalse \@bburyfalse
		\@heightfalse \@widthfalse
		\@rheightfalse \@rwidthfalse
		\def\@p@sbbllx{}\def\@p@sbblly{}
		\def\@p@sbburx{}\def\@p@sbbury{}
		\def\@p@sheight{}\def\@p@swidth{}
		\def\@p@srheight{}\def\@p@srwidth{}
		\def\@p@sfile{}
		\def\@p@scost{10}
		\def\@sc{}
		\@prologfilefalse
		\@postlogfilefalse
		\@clipfalse
		\if@noisy
			\@verbosetrue
		\else
			\@verbosefalse
		\fi
}
\def\parse@ps@parms#1{
	 	\@psdo\@psfiga:=#1\do
		   {\expandafter\@setparms\@psfiga,}}
\newif\ifno@bb
\def\bb@missing{
	\if@verbose{
		\typeout{psfig: searching \@p@sfile \space  for bounding box}
	}\fi
	\no@bbtrue
	\epsf@getbb{\@p@sfile}
        \ifno@bb \else \bb@cull\epsf@llx\epsf@lly\epsf@urx\epsf@ury\fi
}	
\def\bb@cull#1#2#3#4{
	\dimen100=#1 bp\edef\@p@sbbllx{\number\dimen100}
	\dimen100=#2 bp\edef\@p@sbblly{\number\dimen100}
	\dimen100=#3 bp\edef\@p@sbburx{\number\dimen100}
	\dimen100=#4 bp\edef\@p@sbbury{\number\dimen100}
	\no@bbfalse
}
\def\compute@bb{
		\no@bbfalse
		\if@bbllx \else \no@bbtrue \fi
		\if@bblly \else \no@bbtrue \fi
		\if@bburx \else \no@bbtrue \fi
		\if@bbury \else \no@bbtrue \fi
		\ifno@bb \bb@missing \fi
		\ifno@bb \typeout{FATAL ERROR: no bb supplied or found}
			\no-bb-error
		\fi
		\count203=\@p@sbburx
		\count204=\@p@sbbury
		\advance\count203 by -\@p@sbbllx
		\advance\count204 by -\@p@sbblly
		\edef\@bbw{\number\count203}
		\edef\@bbh{\number\count204}
}
\def\in@hundreds#1#2#3{\count240=#2 \count241=#3
		     \count100=\count240	
		     \divide\count100 by \count241
		     \count101=\count100
		     \multiply\count101 by \count241
		     \advance\count240 by -\count101
		     \multiply\count240 by 10
		     \count101=\count240	
		     \divide\count101 by \count241
		     \count102=\count101
		     \multiply\count102 by \count241
		     \advance\count240 by -\count102
		     \multiply\count240 by 10
		     \count102=\count240	
		     \divide\count102 by \count241
		     \count200=#1\count205=0
		     \count201=\count200
			\multiply\count201 by \count100
		 	\advance\count205 by \count201
		     \count201=\count200
			\divide\count201 by 10
			\multiply\count201 by \count101
			\advance\count205 by \count201
		     \count201=\count200
			\divide\count201 by 100
			\multiply\count201 by \count102
			\advance\count205 by \count201
		     \edef\@result{\number\count205}
}
\def\compute@wfromh{
		\in@hundreds{\@p@sheight}{\@bbw}{\@bbh}
		\edef\@p@swidth{\@result}
}
\def\compute@hfromw{
		\in@hundreds{\@p@swidth}{\@bbh}{\@bbw}
		\edef\@p@sheight{\@result}
}
\def\compute@handw{
		\if@height 
			\if@width
			\else
				\compute@wfromh
			\fi
		\else 
			\if@width
				\compute@hfromw
			\else
				\edef\@p@sheight{\@bbh}
				\edef\@p@swidth{\@bbw}
			\fi
		\fi
}
\def\compute@resv{
		\if@rheight \else \edef\@p@srheight{\@p@sheight} \fi
		\if@rwidth \else \edef\@p@srwidth{\@p@swidth} \fi
}
\def\compute@sizes{
	\compute@bb
	\compute@handw
	\compute@resv
}
\def\psfig#1{\vbox {
	\ps@init@parms
	\parse@ps@parms{#1}
	\compute@sizes
	\ifnum\@p@scost<\@psdraft{
		\if@verbose{
			\typeout{psfig: including \@p@sfile \space }
		}\fi
		\special{ps::[begin] 	\@p@swidth \space \@p@sheight \space
				\@p@sbbllx \space \@p@sbblly \space
				\@p@sbburx \space \@p@sbbury \space
				startTexFig \space }
		\if@clip{
			\if@verbose{
				\typeout{(clip)}
			}\fi
			\special{ps:: doclip \space }
		}\fi
		\if@prologfile
		    \special{ps: plotfile \@prologfileval \space } \fi
		\special{ps: plotfile \@p@sfile \space }
		\if@postlogfile
		    \special{ps: plotfile \@postlogfileval \space } \fi
		\special{ps::[end] endTexFig \space }
		\vbox to \@p@srheight true sp{
			\hbox to \@p@srwidth true sp{
				\hss
			}
		\vss
		}
	}\else{
		\if@draftbox{		
			\hbox{\fbox{\vbox to \@p@srheight true sp{
			\vss
			\hbox to \@p@srwidth true sp{ \hss \@p@sfile \hss }
			\vss
			}}}
		}\else{
			\vbox to \@p@srheight true sp{
			\vss
			\hbox to \@p@srwidth true sp{\hss}
			\vss
			}
		}\fi

	}\fi
}}
\def\psglobal{\typeout{psfig: PSGLOBAL is OBSOLETE; use psprint -m instead}}
\psfigRestoreAt

\newcommand{\tJ}{$t$-$J$\ }
\newcommand{\scr}[1]{\mbox{\scriptsize #1}}
\newcommand{\scrscr}[1]{\mbox{\tiny #1}}
\newcommand{\ov}[1]{\overline{#1}}
\newcommand{\dt}{\Delta \tau}
\newcommand{\ww}{\tilde{w}}
\newcommand{\ti}{\tau_{\scr{int}}}
\newcommand{\tio}{\tau^\O_{\scr{int}}}
\newcommand{\TO}{\rightarrow}
\renewcommand{\O}{{\cal O}}
\newcommand{\C}{{\cal C}}
\newcommand{\Cp}{{\cal C}_p}
\newcommand{\G}{{G}}
\newcommand{\Gp}{{G_p}}
\newcommand{\beq}[1]{\begin{equation}\label{#1}}
\newcommand{\eeq}{\end{equation}}
\newcommand{\deffig}[4]{
  \begin{figure}[tbh]
   \begin{center}
    \null\ 
    \psfig{figure=#2,width=#3}
   \end{center}
   \caption[*]{#4}
   \label{#1}
  \end{figure}}
\newcommand{\tabfig}[1]{\parbox{1cm}{\vspace{0.5mm}
                        \psfig{figure=#1,height=.90cm,width=.90cm}}}

\begin{document}

\title{Quantum Monte Carlo Loop Algorithm for the $t$-$J$ Model}
\author{Beat Ammon\cite{ABA}}
\address{SCSC and Theoretische Physik, 
        Eidgen\"ossische Technische Hochschule,
        CH-8092 Z\"urich, Switzerland}
\author{Hans Gerd Evertz\cite{AHGE}}
\address{Theoretische Physik, Univ. W\"urzburg, 97074 W\"urzburg, 
	Germany}
\author{Naoki Kawashima\cite{ANK}}
\address{Department of Physics, Toho University, 
        Miyama 2-2-1, Funabashi, Chiba 274, Japan}
\author{Matthias Troyer\cite{AMT}}
\address{Institute for Solid State Physics, University of Tokyo, 
        Roppongi 7-22-1, Tokyo 106, Japan}
\author{Beat Frischmuth\cite{ABF}}
\address{Theoretische Physik, 
        Eidgen\"ossische Technische Hochschule,
        CH-8093 Z\"urich, Switzerland}


\maketitle

\begin{abstract}
We propose a generalization of the Quantum Monte Carlo loop algorithm
to the $t$-$J$ model by a mapping to three coupled six-vertex
models. The autocorrelation times are reduced by orders of magnitude
compared to the conventional local algorithms.  The method is
completely ergodic and can be formulated directly in continuous time.
We introduce improved estimators for simulations with a local sign
problem.  Some first results of finite temperature simulations are
presented for a $t$-$J$ chain, a frustrated Heisenberg chain, and
$t$-$J$ ladder models.
\end{abstract}

\pacs{71.10.Fd, 71.27+a, 02.70.Lq}

\section{Introduction}
Quantum Monte Carlo (QMC) methods are a powerful tool for the
investigation of strongly interacting systems. They are easy to
generalize and can therefore be applied to almost any model. In
addition, they can be used for large systems and give unbiased results
that are exact within given statistical errors. They are thus an ideal
tool for numerical simulations of complex systems.
A major problem however is that the results are not useful if the
statistical errors become too large. This happens in many interesting
cases. Classical local update Monte Carlo (MC) simulations near second
order phase transitions suffer from ``critical slowing down'': the
autocorrelation time and with it the statistical errors diverge at the
critical point. This problem has been solved for many classical spin
systems by cluster algorithms, which construct global updates of large
clusters instead of performing local spin flips.

Recently a generalization of these cluster methods to quantum spin
systems, the loop algorithm, has been developed.
\cite{hge_prl,hge_more,wiese,Hubbard,nk1,nk2,nk3}  For a review see
Ref.\ \onlinecite{LoopReview}.  This method can solve the problem of
critical slowing down also for QMC simulations. It has made possible
to investigate phase transitions in quantum spin systems,
\cite{Zhang94,CVO,LCO,EXP,Frischmuth96,Greven96,beard,KT,SL,random,beardPT}
far beyond the possibilities of previous MC techniques.

The loop algorithm can be generalized to particle models.  The
original loop method \cite{hge_prl,hge_more} can be applied directly
to hard core bosons and to spinless fermions. A Hubbard model can be
simulated by coupling two spinless fermion systems.\cite{Hubbard} One
problem in QMC simulations of the Hubbard model is that its dominant
energy scale is the Coulomb repulsion $U\gg t$, while the interesting
low lying excitations are at a much smaller energy scale $J=4t^2/U \ll
U$. To investigate the low energy properties it is thus of advantage
to simulate the effective low energy Hamiltonian, the $t$-$J$ model.
Previous finite temperature simulations for the \tJ model have been
carried out both in a determinantal formulation \cite{XYZ} in two
dimensions (2D), which suffered from serious sign problems and
metastability, and in the worldline formulation in one dimension (1D),
with standard MC updates.\cite{Assaad} As we will show
explicitly later, such standard MC simulations suffer from strong
autocorrelations which seriously limit the accessible system sizes and
temperatures.  They are also non-ergodic, and like the determinantal
simulations have to be extrapolated to continuous imaginary time.

In the present paper we present a loop algorithm for the $t$-$J$ model
(for any dimension), which overcomes these autocorrelation problems
and has additional advantages such as complete ergodicity, the
existence of improved estimators, which further reduce the error of
measured quantities by implicitly averaging over many configurations,
and the possibility of directly taking the continuous time limit.
Some of our results have already been presented in Refs.\
\onlinecite{APS-talk,Naoki_Athens}.  An apparently related approach
has very recently been used for 2D simulations in Ref.\
\onlinecite{Brunner}.

Quantum Monte Carlo simulations of fermionic models or of frustrated
spin systems nearly always suffer from the ``negative sign
problem''. In order to perform QMC simulations we first have to map
the quantum system to a classical one. This mapping can introduce
negative weights, which cause cancellation effects. The statistical
error for a given amount of computational effort can then increase
exponentially with system size and inverse temperature. This severely
restricts simulations of higher dimensional fermionic models.  By
combining loop updates with improved estimators we can reduce the
variance of the observables and thus lessen the sign problem.

This paper is organized as follows. First we review the worldline QMC
algorithm and the standard loop algorithm for a Heisenberg chain. In
Sec. \ref{tjloops} we describe the loop algorithm for the $t$-$J$
model. The use of improved estimators is discussed in
Sec. \ref{impr}. Finally in Sec. \ref{results} we discuss the
performance of the new algorithm and show some first results obtained
for a $t$-$J$ chain, a frustrated Heisenberg chain, and $t$-$J$
ladder models. In the appendix we discuss the continuous time version
of our method.

\section{Background Materials} \label{background}
To establish notation and formal background we briefly describe the
worldline representation and the standard loop algorithm.  We refer to
the literature for more detailed descriptions of the worldline
representation \cite{worldlines} and the loop algorithm.
\cite{hge_prl,wiese,nk2,LoopReview} As an example, we take the 1D 
Heisenberg antiferromagnet. The Hamiltonian is defined by
\begin{equation} \label{Eq_Heisenberg}
        H_{H} = \sum_{i=1}^{L} H^{(i)}
                =\sum_{i=1}^{L} J\, \vec{S}_{i}\cdot \vec{S}_{i+1},
\end{equation}
where $\vec{S_{i}}$ denotes a spin-1/2 operator on site $i$, $J>0$ for
the antiferromagnet, and the periodic boundary condition
$\vec{S}_{L+1} = \vec{S}_1$ is adopted.

\subsection{Worldline representation}
We use the Trotter-Suzuki decomposition \cite{trotter_suzuki} and a
path-integral formulation in imaginary time.  The Hamiltonian $H$ is
decomposed into two terms $H=H_{\scr{even}}+H_{\scr{odd}}$, each of
which is easy to diagonalize.  Then
\begin{eqnarray} \label{Trotter-Suzuki}
        Z & = & \mbox{Tr} e^{-\beta H} = 
                \lim_{M\rightarrow\infty}
                \mbox{Tr}\left( \left( e^{-\Delta \tau 
                        \left( H_{\scr{even}}+H_{\scr{odd}} \right)} 
                        \right)^{M} \right) \nonumber 
         =      
                \mbox{Tr}\left( \left( e^{-\Delta \tau H_{\scr{even}}} 
                       e^{-\Delta \tau H_{\scr{odd}}}\right)^{M}\right) 
                        + O \left(\Delta \tau ^{2} \right)   \\
        & = &  
               \sum_{i_{1},\ldots ,i_{2M}}
               \langle i_{1}  | e^{-\Delta \tau H_{\scr{even}}} | 
			i_{2M}\rangle
               \langle i_{2M} | e^{-\Delta \tau H_{\scr{odd}}} | 
			i_{2M-1}
							\rangle \cdots 
               \langle i_{3}  | e^{-\Delta \tau H_{\scr{even}}} | 
			i_{2}\rangle 
               \langle i_{2}  | e^{-\Delta \tau H_{\scr{odd}}} | 
			i_{1}\rangle 
                        + O \left(\Delta \tau ^{2} \right), 
\end{eqnarray}
where $\Delta \tau=\beta / M$, and $M$ is called Trotter-number. The
summation with respect to $| i_{k} \rangle$ is taken over  complete
orthonormal sets of states.

We may consider Eq.~(\ref{Trotter-Suzuki}) as the evolution of
the initial state $| i_{1} \rangle$ in imaginary time 
with one application of the time evolution operator  
within a time step $\Delta\tau$. 
The partition function $Z$ in Eq.~(\ref{Trotter-Suzuki}) is also
formally the partition function of a (d+1) dimensional classical
system.
The systematic error of order $\Delta \tau^{2}$ due to the finite time
step approximation can be extrapolated to $\Delta \tau^{2} \rightarrow
0$ by fitting to a polynomial in $\Delta \tau^{2}$.  The loop
algorithm can also be formulated directly in the continuous time limit
$\Delta \tau\rightarrow 0$.\cite{beard} This will be discussed 
in appendix \ref{continuous}.  
\deffig{fig_worldline}{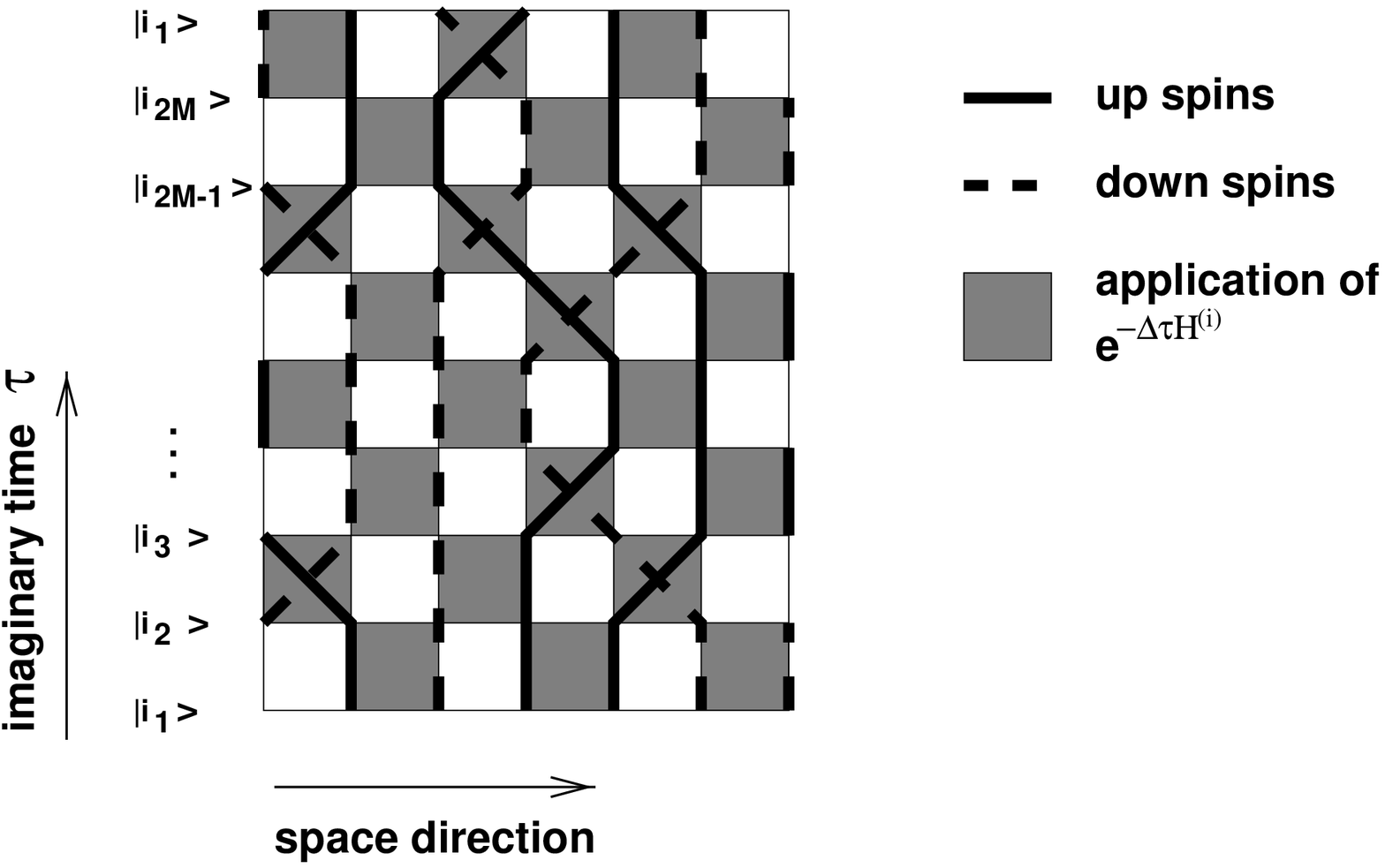}{80mm}
{Example of a world line configuration of the Heisenberg model with
the checkerboard decomposition. The imaginary time $\tau$ runs along
the vertical axis, and the real space direction along the horizontal
axis. The solid lines represent up-spins, the dashed lines down
spins. The shaded plaquettes show the application of $e^{-\dt
H^{(i)}}$.}

The decomposition of the Hamiltonian has to be chosen according to the
problem.  For our 1D system with only nearest neighbor interaction we
take $H_{\scr{even/odd}}=\sum_{i \mbox{ }\scr{even/odd}} H^{(i)}$,
leading to a checkerboard structure as shown in
Fig.~\ref{fig_worldline}.  The Hamiltonian acts only on the shaded
plaquettes $p$ in Fig.~\ref{fig_worldline}, each of which contributes
a factor $w_p$ to the matrix elements in Eq.~(\ref{Trotter-Suzuki}),
i.e.,
\begin{equation} \label{weights}
  Z= \sum_{i_{1},\ldots ,i_{2M}}  
  \prod_{p} w_p \equiv \sum_{\{{\cal C}\}}W({\cal C})\,.
\end{equation}

\begin{figure}[tb]
\begin{center}
\begin{tabular}{|c@{\hskip1ex}|cc|cc|cc|}
\hline
  $\Cp$  &  
            \quad \tabfig{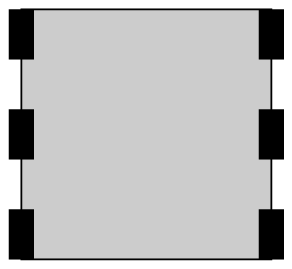} \quad & \quad \tabfig{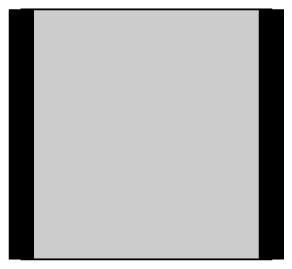}
		 \quad &
            \quad \tabfig{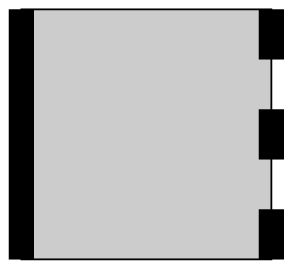}  \quad & \quad \tabfig{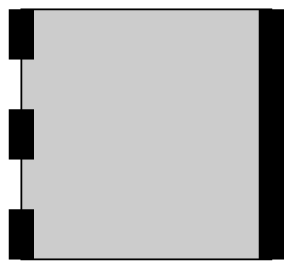}
		 \quad &
            \quad \tabfig{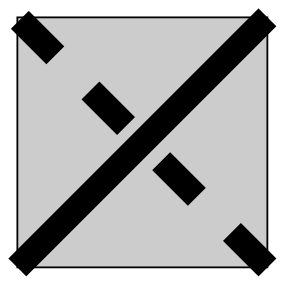} \quad & \quad \tabfig{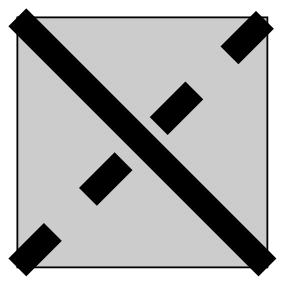} 
		 \quad 
\\[2mm]
\hline
 & & & & & & \\[-2mm]
  $w(\Cp)$ & 
         \multicolumn{2}{c|}{$e^{-\Delta\tau J/4}$} &
         \multicolumn{2}{c|}{$e^{\Delta\tau 
		J/4}{\rm ch}(\Delta\tau J/2)$} &
         \multicolumn{2}{c|}{$e^{\Delta\tau 
		J/4}{\rm sh}(\Delta\tau J/2)$} 
\\[2mm]
\hline
\end{tabular}
\end{center}
\caption[*]{
  The six allowed plaquette states of the Heisenberg model that
  fulfill the magnetization-conservation condition.  The second row
  shows the weights of the plaquettes for the Hamiltonian
  (Eq.~\ref{Eq_Heisenberg}).  The solid lines connect two sites
  occupied by up spins, and the dashed ones connect down spins.  (We
  have assumed a bipartite lattice and rotated the spin-operators
  $S^{x,y} \rightarrow - S^{x,y}$ to make the weight of the last two
  plaquettes positive.)
\vspace*{0ex}  }
\label{fig_HB}
\end{figure}

For the Heisenberg model we expand the states $| i_{k} \rangle$ in an
$S^z$ eigenbase.  Each $H^{(i)}=\vec{S}_i \cdot \vec{S}_{i+1}$
conserves magnetization.  Therefore there are only six nonvanishing
matrix elements for each shaded plaquette, which can be represented by
solid and dashed lines connecting up and down spins, respectively, as
shown in Fig.~\ref{fig_HB}.  Thus the sum in Eq.~(\ref{weights}) is
taken over configurations $\C \equiv \{ |i_{k}\rangle \}$ of {\em
continuous worldlines}.  One example is shown in
Fig.~\ref{fig_worldline}.  Note that $\C$ can be identified with a set
of binary variables $S^z_i=\pm 1/2$, each defined on a site in the
checkerboard, with the restriction due to the magnetization
conservation.  It is convenient to define $\Cp$ as the local state of
a given shaded plaquette $p$, namely, a set of four binary variables
at its four corners.  The configuration $\C$ is then identified with
the union of the plaquette states.  Accordingly, $w_p$ in
Eq.~(\ref{weights}) can be written as $w(\Cp)$.

Thermal averages of observables $\cal O$ can be written in a similar
way as
\begin{equation}
\langle {\cal O} \rangle = {1\over Z}\sum_{\{{\cal C}\}}
	W({\cal C}) {\cal O}({\cal C})\,,
\label{eq:meas}
\end{equation}
where ${\cal O}({\cal C})$ is the value of the observable in the
configuration $\cal C$.

If the weight of a configuration $W(\C)$ can take negative values, one
has to use its absolute value $|W(\C)|$ to construct the probabilities
for the Markov chain of a MC procedure (see below), since
these probabilities need to be positive.
Expectation values are then given by
\begin{eqnarray}\label{eq:meas_sign}
\langle {\cal O} \rangle
&=& \frac
 {\sum_{\{{\cal C}\}} W({\cal C}) \,              {\cal O}({\cal C})}
 {Z} 
\;=\;\frac
 {\sum_{\{{\cal C}\}}|W({\cal C})|\,\mbox{sign}({\cal C}) 
		\,{\cal O}({\cal C})}
 {\sum_{\{{\cal C}\}}|W({\cal C})|\,\mbox{sign}({\cal C})} \nonumber \\
&=& \frac{\frac
 {\sum_{\{{\cal C}\}}|W({\cal C})|\,\mbox{sign}({\cal C}) 
		\,{\cal O}({\cal C})}
 {\sum_{\{{\cal C}\}}|W({\cal C})|}}
 {\frac
 {\sum_{\{{\cal C}\}}|W({\cal C})|\,\mbox{sign}({\cal C})} 
 {\sum_{\{{\cal C}\}}|W({\cal C})|}}
\;=\;\frac
 {\langle \mbox{sign} \cdot {\cal O} \rangle_{|W|}}
 { \langle \mbox{sign} \rangle_{|W|}} \;,
\label{eq:smeas}
\end{eqnarray}
where $sign(\C)$ stands for the sign of $W(\C)$, and
$\langle \cdots \rangle_{|W|}$ denotes expectation values
with respect to the absolute value of the weight $W$.

In many cases, a ``sign problem'' now stems from the fact that 
the average sign,  $\langle \mbox{sign} \rangle_{|W|}$,
may decay exponentially with increasing 
system size and inverse temperature $\beta$. 
For fixed computational effort this then leads to an
exponential blow-up of the errors.

\subsection{Local worldline algorithms}

The thermal averages Eqns.~(\ref{eq:meas}, \ref{eq:meas_sign}) can be
taken by MC importance sampling.  One constructs a sequence (Markov
chain) of configurations $\C^{(i)}$ such that in the limit of
infinitely many configurations their distribution agrees with the
correct Boltzmann distribution $p(\C^{(i)})=W(\C^{(i)})/Z$.

This can be achieved by satisfying two conditions: 
ergodicity of the Markov chain, 
and detailed balance 
\begin{equation} \label{detbal}
  \frac{p(\C \rightarrow \C')}{p(\C' \rightarrow \C)}
      \;=\;  \frac{W(\C')}{W(\C)} \,,
\end{equation} 
where $p(\C \rightarrow \C')$ is the probability
of choosing the configuration $\C'$ as the next configuration
in the Markov chain, when the current configuration is $\C$.

Then the thermal expectation value Eq.~(\ref{eq:meas}) of an
observable $\cal O$ can be estimated by averaging the
value of the observable in the configurations $\C^{(i)}$: 
\begin{equation}
\label{observables}
        \langle {\cal O}\rangle = \lim_{N\TO\infty} 
		\ov{\cal O} \;,\;\;\;
                    \ov{\cal O} = {1\over N}
			\sum_{i}^N {\cal O}(\C^{(i)}).
\end{equation}
In cases with a sign problem the averages in Eq.~(\ref{eq:meas_sign})
are done separately for the numerator and for the denominator:
\begin{equation}
\label{Osign}
\ov{\cal O} =\frac
 {{1\over N}\sum_{i=1}^N (\mbox{sign} \cdot {\cal O})_{|W|}^{(i)}}
 {{1\over N}\sum_{i=N}^n (\mbox{sign})_{|W|}^{(i)}}.
\end{equation}

In standard local algorithms an update from one configuration $\C$ to
the next one is done by proposing a new configuration $\C'$ that
differs from $\C$ by a small {\em local} change of the worldlines.
The candidate $\C'$ is accepted with a probability that
satisfies detailed balance, e.g.\ the Metropolis
probability \cite{metropolis}
\begin{equation}\label{p_metrop}
  p(\C\rightarrow \C') \,=\, 
  \min\left( 1,\frac{W(\C')}{W(\C )}\right),
\end{equation}
or the heatbath probability
\begin{equation} \label{p_symm}
  p(\C\rightarrow \C') \,=\, 
          \frac{W(\C')}{W(\C)+W(\C')}\ \;\;;
\end{equation}
otherwise the configuration $\C$ is kept.

There are two major problems with local updates: Firstly, consecutive
configurations are strongly correlated.  It takes on average a number
$\tau$ of updates to arrive at a statistically independent
configuration.  This {\em autocorrelation time} $\tau$, which depends
on the measured quantity ${\cal O}$, typically increases quadratically
with spatial correlation length $\xi$ and inverse energy gap
$\Delta^{-1}$ (resp.\ system size $L$ and inverse temperature $\beta$
when $\xi > L$ or $\Delta^{-1} > \beta$).  To achieve a desired
statistical accuracy, the MC simulation has to be lengthened by a
factor $\tau$, which can easily reach orders of $10^6$ and larger in
practical cases.  Secondly, local updates are {\em not ergodic}.  For
example, when applied to the Heisenberg model, they cannot change
total magnetization nor spatial winding number.  Many quantities of
physical interest, like the superfluid density, are then very
difficult to estimate.  In addition, it was pointed out \cite{nk2}
that a complicated quantity exists that does not vary in conventional
local updates for the $XYZ$ model.  To make the conventional algorithm
ergodic, therefore, we usually have to include some ad hoc global
updates, which tends to make the resulting code rather cumbersome.
Also, the acceptance rate of such ad hoc global updates is often very
small, which is another cause of long autocorrelation times.

\subsection{Loop algorithm for the Heisenberg model}

Both kinds of difficulties are overcome in the loop algorithm which
achieves large nonlocal configuration changes in one stochastic
update.  Autocorrelation times for the loop algorithm are found to be
orders of magnitude smaller than those for the conventional
algorithm. In addition, it does not suffer from the
above-mentioned ergodicity problems,
and can be formulated directly in continuous time.\cite{beard}

In the loop algorithm, each update consists of two steps, both
stochastic.  In the first step, the current worldline configuration
$\C$ is mapped with a probability $p(\C \TO \G)$ to a
graph-configuration $G$.  $\G=\{\Gp\}$ consists of local graph
segments $\Gp$ defined on the plaquettes $p$, which combine to form a
set of closed loops.  In the second step, the configuration of loops
is mapped with a probability $p(\G \TO \C')$ to a new worldline
configuration $\C'$.

Let us explain the simple case of the Heisenberg model.  Two
observations are important: (1) Since the worldlines are continuous,
the difference between two arbitrary worldline configurations (in the
sense of an exclusive-or), i.e.\ the {\em location of spin-flips} in
any allowed update of a configuration, is located on a set of closed
loops. These are the loops we will construct. Flipping all the spins
on a closed loop will be called a loop flip. In Fig.~\ref{fig_updates}
we show an example for a loop-update step for the Heisenberg model.
(2) Since the Hamiltonian acts locally, the partition function $Z$ in
Eq.~(\ref{weights}) is a product of plaquette terms. We can therefore
fulfill detailed balance separately for each plaquette, provided the
global constraint of closed loops is satisfied.
\deffig{fig_updates}{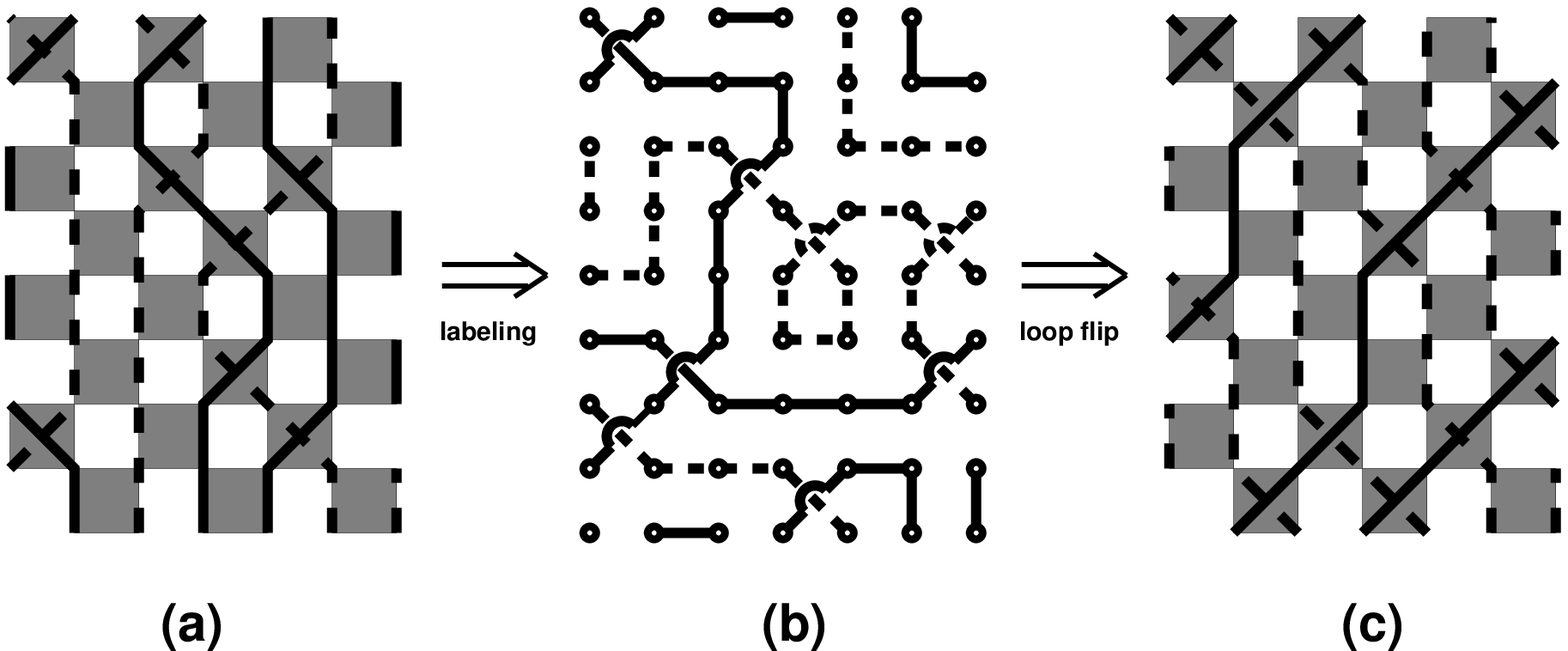}{120mm}
{Example of a loop-update step for the Heisenberg model (anisotropic
case, which has finite probability for diagonal graph segments).  On
the left (a) we show an initial configuration $\C$ of up-spins (dashed
lines) and down-spins (solid lines) which is mapped to a configuration
$\G$ of loops in the middle figure (b).  Some of these loops are
selected (with probability 1/2) to be flipped, i.e.\ the spins along
these loops change direction. We denote the loops that will be flipped
by dashed lines, the unchanged loops by solid lines.  The figure on the
right (c) shows the spin configuration after the loop-flips.}

\begin{figure}[tb]
 \begin{center}
  \begin{tabular}{|c|c|c||c|c|c|c|} \hline
     \multicolumn{3}{|r||}{$\Gp$ } &
        \tabfig{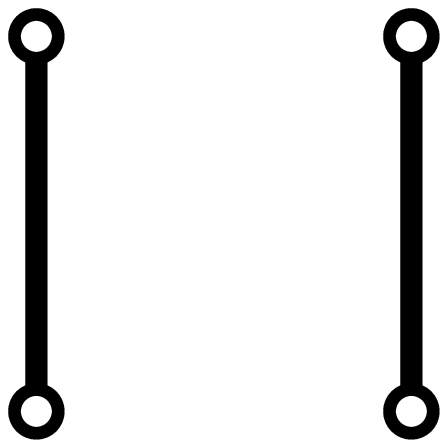} & 
        \tabfig{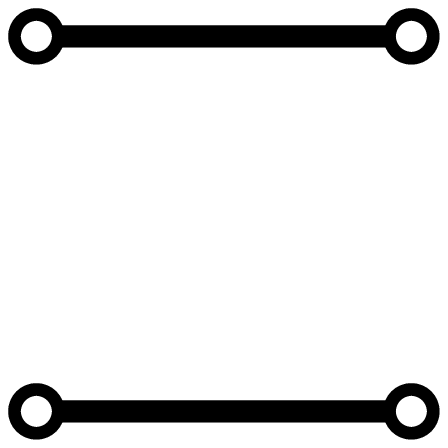} & 
        \tabfig{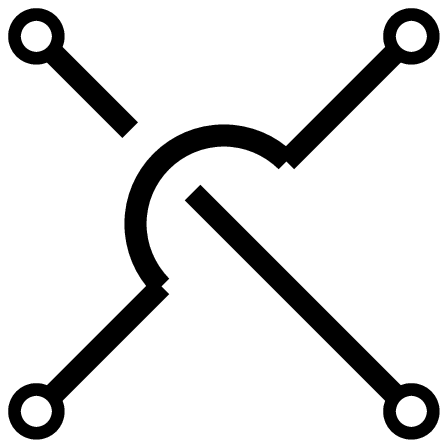} &
        \tabfig{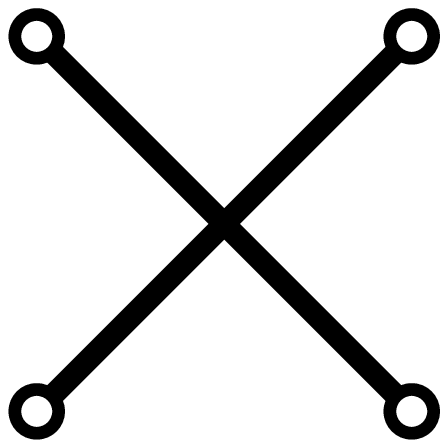} \\\hline
     \multicolumn{3}{|r||}{$v(\Gp)$} & 
        $e^{-\Delta \tau J/4} - \epsilon J \Delta\tau/2$ &
        $e^{\Delta \tau J/4} \mbox{sh}(\Delta\tau J/2)-\epsilon 
		J\Delta\tau/2$&
        $\epsilon J \Delta\tau /2$ & 
        $\epsilon J \Delta\tau$ \\\hline
     \multicolumn{3}{|r||}{$\lim_{\Delta\tau\rightarrow
		 {\rm d}\tau}v(\Gp)$} & 
        $1$ &
        $(1-\epsilon)(J/2){\rm d\tau}$ & 
        $(J\epsilon/2){\rm d}\tau$ & 
        $J \epsilon{\rm d}\tau$ \\\hline\hline
     $\Cp$ & $w_{0}(\Cp)$ & {$w_{\scr{asymm}}(\Cp)$} & 
        \multicolumn{4}{c|}{$\Delta(\Gp,\Cp)$} \\\hline\hline
%
     $\tabfig{phi2b.eps} \tabfig{phi2.eps}$ & 
        $e^{-\Delta \tau J/4}$ & 
        1 & 
        1 & 0 & 1 & 0 \\
     $\tabfig{phi3.eps} \tabfig{phi3b.eps}$ &
        $e^{\Delta \tau J/4} \mbox{ch}(\Delta \tau J/2)$ & 
        1 & 
        1 & 1 & 0 & 1 \\
     $\tabfig{phi1b.eps} \tabfig{phi1.eps}$ & 
        $e^{\Delta \tau J/4} \mbox{sh}(\Delta \tau J/2)$ & 
        1  & 
        0  & 1 & 1 & 0 \\
\hline
\end{tabular}
\vspace*{2ex}
\caption[*]{
        Plaquette configurations $\Cp$ and graphs $\Gp$ for the
        anti-ferromagnetic Heisenberg model.  The upper part of the
        figure specifies the graphs $\Gp$ and one solution $v(\Gp)$ of
        equation (\ref{weight_equation}) for their weights.  There is
        a free parameter $\epsilon$ in this solution.  If $\epsilon$
        is chosen zero, no freezing and no diagonal graph segments
        will occur.  The third row shows the the continuous time limit
        of $v(\Gp)$ (see appendix \ref{continuous}).  The lower part
        of the figure shows the spin configurations $\Cp$ and their
        weights, and the function $\Delta(\Gp,\Cp)$ which specifies
        whether a configuration $\Cp$ and a graph $\Gp$ are
        compatible.}  \label{FigHB_labels}
\end{center}
\end{figure}

By inspecting the six allowed local states $\Cp$ on a plaquette
(Fig.~\ref{fig_HB}), we see that for each plaquette, spin-flips must
occur on {\em pairs} of sites, not on single spins, in order to arrive
at another allowed local state.  We connect the pairs of sites 
on which spins are to be flipped together by
solid lines: these are loop segments.  Since there are several
possible pairings of sites, the lines can in principle run
horizontally, vertically, or diagonally.  Also, all four sites can be
flipped simultaneously without violating the restriction.
The symbol $\Gp$ stands for the two loop segments on plaquette $p$.
We will speak of $\Gp$ as the graph on plaquette $p$.  The union
$\G=\cup_p \Gp$ constitutes a complete graph configuration $\G$.  We
call the graph $\Gp$ in which all 4 spins are grouped together the
``freezing'' graph, since (without symmetry breaking field) the flip
of all 4 spins will leave the plaquette weight $w_p$ invariant.
For a given plaquette configuration $\Cp$ only certain graphs $\Gp$
are possible, namely those for which an update along connected points
leads to another allowed plaquette configuration $\C_p'$ 
(i.e.\ $w(\C_p') \ne 0$).
We define a function $\Delta(\Gp,\Cp)$ so that it takes the value 1
when $\Gp$ is allowed for a given $\Cp$ and the value 0 otherwise.  It
is shown in Fig.~\ref{FigHB_labels}.  Each spin belongs to two
interacting plaquettes.  It belongs to one loop segment on each of
these plaquettes, except for the ``freezing'' graph, in which all four
sites are connected.
Specifying $\Gp$ on each interacting plaquette individually will thus
automatically lead to an overall configuration of closed loops.
Therefore, we only need to specify the probabilities $p(\Cp \TO \Gp)$
and $p(\Gp \TO \Cp')$ for each interacting plaquette individually.
If we additionally have ``freezing'' graphs $\Gp$ on some of the
plaquettes, then the loops passing through these plaquettes have to be
flipped together. They are ``frozen'' into one loop-cluster. This
freezing is problematic if it occurs too often since then the whole
lattice might just ``freeze'' and no change of weight is done by the
update. Thus we want to avoid unnecessary freezing.

We may construct loop algorithms such that loops are flipped with
probability $1/2$ when there is no symmetry breaking field, as is the
case with general $XYZ$ quantum spin systems without magnetic field.
To include a symmetry breaking field,
we factorize the local Boltzmann weight in the form
\begin{equation}\label{factorize}
  |w(\Cp)| = w_0(\Cp) \, w_{\scr{asymm}}(\Cp)
\end{equation}
where $w_0(\Cp)$ is used for defining the probability of choosing
$\Gp$, whereas $w_{\scr{asymm}}(\Cp)$ is taken into account in terms
of the flipping probability of the loop. The weight $w_0(\Cp)$ needs
to be invariant \cite{LoopReview} under flip of all four spins at the
plaquette $p$.  Using this factorization, the probability $p(\Cp \TO
\Gp)$ is constructed as follows.  First we {\em choose} weights
$v(\Gp)$ for all graphs $\Gp$ such that
\begin{equation} 
  \label{weight_equation}
  \sum_\Gp v(\Gp) \, \Delta(\Gp,\Cp) \,=\, w_0(\Cp) .
\end{equation}
One solution to this set of equations is shown in
Fig.~\ref{FigHB_labels}.  (The solution is in general not unique;
depending on $H$, it may also not exist.)  Then, detailed balance for
the overall update $\C \TO \C'$ is fulfilled by
\begin{equation} \label{labeling_probabilities}
     p(\Cp \TO \Gp) \,=\, \frac{v(\Gp)\,\Delta(\Gp,\Cp)}{w_0(\Cp)}
     \qquad ; \qquad
     p(\G \TO \C') \,=\,\frac{\prod_p w_{\scr{asymm}}(\Cp') 
		\Delta(\Gp,\Cp')}
                             {\prod_p w_{\scr{asymm}}(\Cp) + 
				\prod_p w_{\scr{asymm}}(\Cp')}
                              \,,
\end{equation}
as can be checked easily.
In general, $p(\G \TO \C')$ needs to satisfy detailed balance 
with respect to the weight $\prod_p w_{\scr{asymm}}(\Cp') 
\Delta(\Gp,\Cp')$. 
Here we have chosen a heatbath probability like Eq.~(\ref{p_symm}).
%

The construction of the loops can be performed in a multi-cluster
scheme. In this case, a graph $\Gp$ is chosen on all plaquettes $p$,
and we obtain a unique partitioning of the lattice into $n_{l}$ loops
$l_{i}$. Then we attempt to flip all loops $l_{i}$ according to
Eq.~(\ref{labeling_probabilities}).
We can also use a single-cluster variant.\cite{wolff} We can think of
this variant as picking a single cluster $l_{i}$ of the above
partitioning with a probability $p(|l_{i}|)$ according to the size of
the loop $|l_{i}|=\sum_{\scr{site }(\tau,j)
\in l_{i}} 1$. This loop is then flipped with respect to 
weights $\ww=\prod_{p\in l_i}w_{\scr{asymm}}(\C_p)$ and
Eq.~(\ref{p_metrop}), which means that we will always flip that loop
if $\ww=1$ for all plaquettes. In an implementation of this algorithm
we construct this single loop by picking randomly any site of the
lattice and building a single loop by choosing graphs $\Gp$ only on
the plaquettes along the path.  Hence we need an effort only
proportional to the length $|l_{i}|$.

\section{Loop algorithm for the \mbox{$\lowercase{t}$-$J$} model}
\label{tjloops}
\begin{figure}[tb]
\begin{center}
\begin{tabular}{|c@{\hskip1ex}|@{\hskip1ex}l|}
   \hline
        $w(\Cp)$ & \multicolumn{1}{c|}{$\Cp$} 
        \\
   \hline
        1
                & \tabfig{phi2.eps} \tabfig{phi2b.eps}  
		\tabfig{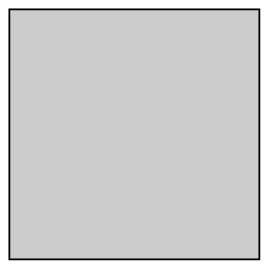}
        \\
        $e^{\dt J/2}{\rm ch} (\dt J/2)$ 
                & \tabfig{phi3.eps} \tabfig{phi3b.eps}
        \\
        $-e^{\dt J/2}{\rm sh} (\dt J/2)$ 
                & \tabfig{phi1.eps} \tabfig{phi1b.eps}
        \\
        ${\rm ch}(\dt t)$ 
                & \tabfig{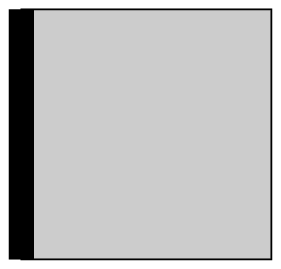}  \tabfig{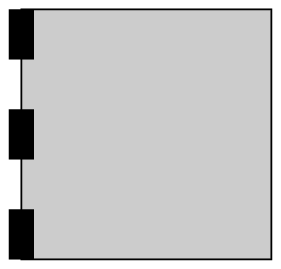} 
                  \tabfig{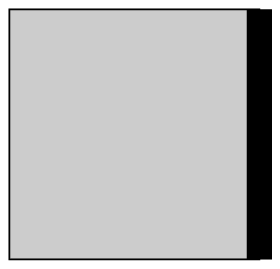} \tabfig{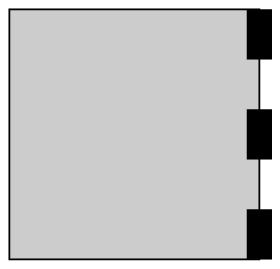}
        \\
        ${\rm sh}(\dt t)$ 
                & \tabfig{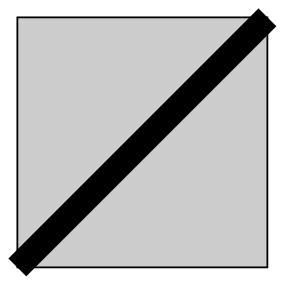}  \tabfig{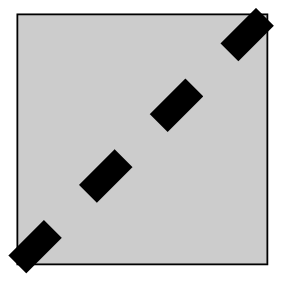} 
                  \tabfig{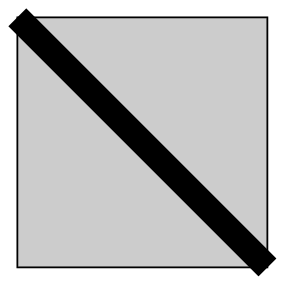} \tabfig{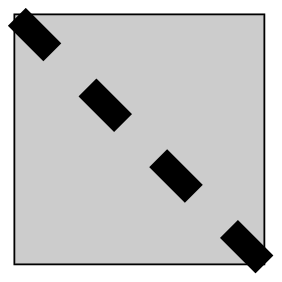}\\
   \hline
\end{tabular}   
\end{center}
\caption[*]{The 15 different plaquettes $\Cp$ with nonvanishing weight
	$w(\Cp)$ for the $t$-$J$ model. Up spins are denoted by a
	solid line, down spins by a broken line.  The sign of $w(\Cp)$
	will be taken into account according to Eq.~(\ref{Osign}).  }%
\label{tJ_weights}
\end{figure}

The $t$-$J$ model is defined by the Hamiltonian
\begin{equation}
  H = -t \sum_{<i,j>}\sum_{\sigma} \left[(1-n_{j,-\sigma})
        c^{\dagger}_{j,\sigma} c_{i,\sigma} (1-n_{i,-\sigma})
        +h.c.\right]
         +J \sum_{<i,j>} (\vec{S}_i \vec{S}_j-{1\over4}n_in_j) \,,
\end{equation}
where $c^{\dagger}_{i,\sigma}$ creates a spin-1/2 fermion with
$z$-component of spin $\sigma$ at site $i$, $n_{i,\sigma}$ =
$c^{\dagger}_{i,\sigma}c_{i,\sigma}$ and
$n_i=\sum_{\sigma}n_{i,\sigma}$. The projection operators
$(1-n_{j,-\sigma})$ prohibit double occupancy of a site. The brackets
$<i,j>$ denote nearest neighbor pairs.
The \tJ model can be represented in a worldline formulation
\cite{Assaad} in terms of variables which take three possible values,
$0$, $+1$ and $-1$, representing a hole, an up-spin, and a down-spin,
respectively. The matrix elements for the 15 different plaquettes
with nonzero weight are given in Fig.~\ref{tJ_weights}.
There are several sources of negative signs in the overall weight
$W(\C)=|W(\C)|\,\mbox{sign}(\C)$ of a configuration. For the \tJ\
model they all stem from anticommutation of fermion operators. One
example is the sign in the third line of Fig.\ \ref{tJ_weights}. The
overall sign can be decomposed as
\begin{equation}\label{sign}
 \mbox{sign}(\C) = (-1)^{n_{\scrscr{perm}}} \, 
			(b_c)^{n_{\scrscr{bound}}} \;,
\end{equation}
where $n_{\scr{perm}}$ is the number of permutations of fermion
worldlines, $b_{c}=+1$ for periodic and $b_{c}=-1$ for antiperiodic
boundary conditions, and $n_{\scr{bound}}$ is the number of particles
hopping across the boundary.
For constructing loops we will use the absolute value of the weight,
Eq.~(\ref{factorize}). The sign will be taken into account in the MC
simulation according to Eq.~(\ref{Osign}).  It will also play a role
for the improved estimators treated in section \ref{impr}.

In the last section, we have seen how a loop algorithm is constructed
for a model with binary variables.  In order to construct a loop
algorithm for the $t$-$J$ model we now reduce the problem with
trivariate variables into three sub-problems with binary variables. To
this end, we divide a MC step into three substeps. In substep
I, variables with the value 0, namely holes, are left unaffected
(inactive) while attempts are made to flip all the variables with
values $+1$ and $-1$ (active variables).  Similarly, in the second and
the third substeps, we keep variables with the values $+1$ and $-1$,
respectively, unaffected.  Therefore, we deal with a binary problem in
each substep.  To each of these binary problems, we apply the idea of
the loop algorithm.  We denote as ``active plaquettes'' those on which
all 4 variables are active.  On the active plaquettes, the resulting
algorithm for substep I is identical to the loop algorithm for the
$S=1/2$ antiferromagnetic Heisenberg model, while the algorithm for
substeps II and III turns out to be the loop algorithm for the $S=1/2$
$XY$ model (which is the same as that for free fermions), as we will
see below. The flipping probabilities of the loops are of course 
affected by the inactive plaquettes.

Since we have three different binary problems, we need to construct
three loop algorithms with the second and the third ones being
transformable into one another simply by interchanging the roles of
the values $+1$ and $-1$. The detailed balance condition holds for
each of the three substeps, whereas ergodicity is achieved by the
combination of them.  We have ample freedom in choosing a set of
graphs and graph weights.  It is, however, advantageous for the
computational simplicity and the reduction of autocorrelation times to
choose a scheme such that the resulting loops may be flipped
independently in a multi-cluster variant (for a different choice see
Ref.\ \onlinecite{Naoki_Athens}).  Therefore, we must have weights
$w_{\scr{asymm}}(\Cp) \equiv 1$ on the active plaquettes, i.e.\ those
where two loops may be flipped.  This can be achieved by letting the
loop updates deal only with the active plaquettes.  The weights of the
other plaquettes are put into the global weight function
$w_{\scr{asymm}}$:
\begin{eqnarray}
  w_{0}(\Cp) & =& \left\{ 
    \begin{array}{ll}
       w(\Cp)  & \mbox{if all four variables on the plaquette 
			are active}\\
       1,      & \mbox{otherwise}
    \end{array}
  \right., \nonumber \\
  w_{\scr{asymm}}(\Cp) & =& \left\{ 
    \begin{array}{ll}
       1,      & \mbox{if all four variables on the plaquette 
			are active}\\
       w(\Cp), & \mbox{otherwise}
    \end{array}
  \right. .
\label{eq:WeightDecomposition}
\end{eqnarray}
In this case we can flip all loops independently with the flipping
probability for a loop $l_{i}$ 
\begin{equation} \label{flip_tJ}
  p_{\scr{flip}}(l_{i})
    =\frac{         
      \prod_{p \in \scr{loop }l_{i}} w_{\scr{asymm}}(\Cp')
     }
     {
      \prod_{p \in \scr{loop }l_{i}} w_{\scr{asymm}}(\Cp)
      +
      \prod_{p \in \scr{loop }l_{i}} w_{\scr{asymm}}(\Cp')
     },
\end{equation}
where $\Cp'$ denotes the plaquette state after the flipping.

\begin{figure}[tb]
 \begin{center}
  \begin{tabular}{|c|c|c||c|c|c||c|c|c|} \hline
     \multicolumn{3}{|r||}{$\Gp$ } &
       \tabfig{psia.eps} & 
       \tabfig{psib.eps} & 
       \tabfig{psic.eps} & 
       \tabfig{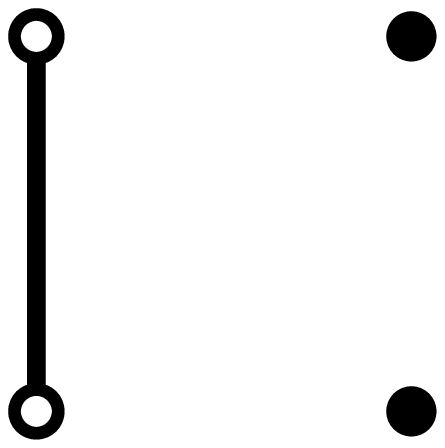} & 
       \tabfig{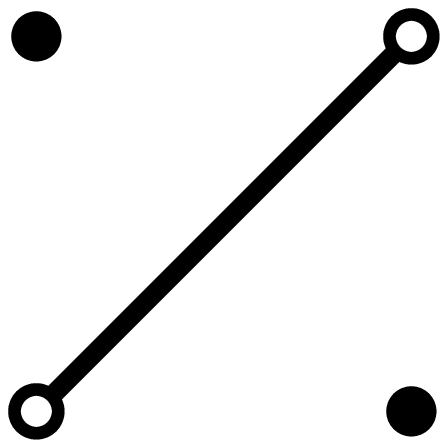} & 
       \tabfig{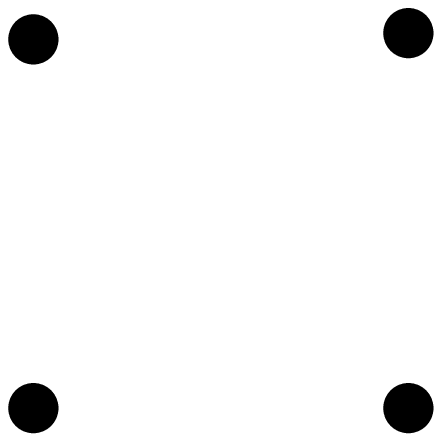} \\\hline
     \multicolumn{3}{|r||}{$v(\Gp)$} & 
        1 & 
        $(e^{\dt J}-1)/2$ & 
        0 & 
        1 & 
        1 & 
        1 \\\hline
     \multicolumn{3}{|r||}{$\lim_{\Delta\tau\rightarrow 
		{\rm d}\tau}v(\Gp)$} & 
        1 & 
        $(J/2){\rm d}\tau$ & 
        0 & 
        1 & 
        1 & 
        1 \\\hline\hline
%
     $\Cp$ & $w_{0}(\Cp)$ & {$w_{\scr{asymm}}(\Cp)$} & 
        \multicolumn{6}{c|}{$\Delta(\Gp,\Cp)$} \\\hline\hline
     \tabfig{phi2b.eps} \tabfig{phi2.eps}  & 
     1 & 
     1 &
     1 & 0 & 1 & 0 & 0 & 0 \\ 
     \tabfig{phi3.eps} \tabfig{phi3b.eps} & 
     $e^{\Delta \tau J/2} {\rm ch}(\Delta \tau J/2)$ &
     1 &
     1 & 1 & 0 & 0 & 0 & 0 \\
     \tabfig{phi1b.eps} \tabfig{phi1.eps} & 
     $e^{\Delta \tau J/2} {\rm sh}(\Delta \tau J/2)$ & 
     1 &
     0 & 1 & 1 & 0 & 0 & 0 \\ 
\hline
     \tabfig{phi4.eps} \tabfig{phi4p.eps} 
     (\hspace*{-1mm} \tabfig{phi4b.eps} 
		\tabfig{phi4bp.eps}\hspace*{-0mm})  & 
     1 & 
     ${\rm ch}(\dt t)$ &
     0 & 0 & 0 & 1 & 0 & 0 \\
     \tabfig{phi5.eps} \tabfig{phi5p.eps} 
     (\hspace*{-0mm}\tabfig{phi5b.eps} 
		\tabfig{phi5bp.eps}\hspace*{-0mm})  & 
     1 & 
     ${\rm sh}(\dt t)$ &
     0 & 0 & 0 & 0 & 1 & 0 \\
     \tabfig{phi2p.eps} & 
     1 & 
     1 &
     0 & 0 & 0 & 0 & 0 & 1 \\\hline
\end{tabular}
\vspace*{2ex}
\caption[*]{
        Plaquette configurations $\Cp$ and graphs $\Gp$ for substep I
        (flip spin-up $\leftrightarrow$ spin-down) of the $t$-$J$
        model.  The upper part of the figure specifies the graphs
        $\Gp$ and the freezing-less solution $v(\Gp)$ of equation
        (\ref{weight_equation}).  The lower part of the figure shows
        the spin configurations $\Cp$ and their weights, and the
        function $\Delta(\Gp,\Cp)$ which specifies whether a
        configuration $\Cp$ and a graph $\Gp$ are compatible.  The
        first six configurations $\Cp$ and the solution $v(\Gp)$
        restricted to these configurations correspond to the case of
        the antiferromagnetic Heisenberg model.  The open circles in
        the diagrams in the top row represent active variables whereas
        solid circles stand for inactive ones.  (For the plaquette
        configurations $\Cp$ in brackets, the corresponding graph
        $\Gp$ given in the figure has to be flipped spatially.)  }
        \label{tJ_labels1}
\end{center}
\end{figure}

Let us consider now in detail the probabilities in the algorithm for
substep I in which variables with the value 0 (holes) are inactive and
kept unaffected.  The algorithm is equivalent to the one for the
$S=1/2$ antiferromagnetic isotropic Heisenberg model as far as the
plaquettes with only active variables are concerned.  As for the
plaquettes with inactive variables, a unique graph is assigned to each
of them such that active variables, if any, are connected to each
other (see lower part of Fig.~\ref{tJ_labels1}).
It is easy to verify that the graph weights $v(\Gp)$ shown in
Fig.~\ref{tJ_labels1} satisfy the weight equation
(\ref{weight_equation}).  In this algorithm, the weights
$w_{\scr{asymm}}(\Cp)$ remain unchanged upon flipping of a loop
because of the spin-inversion symmetry of the Hamiltonian.  Thus we
obtain a loop flipping probability of 1/2.

%

Next, we consider the algorithm for substep II
 (or equivalently substep III), 
where all down-spins are kept unchanged.  
This time, the algorithm on the active plaquettes is equivalent 
to the one for the $S=1/2$ $XY$ model (since that is the same 
as the algorithm for free fermions \cite{LoopReview})
rather than that for the antiferromagnetic Heisenberg model.  Again, a
unique graph is assigned to each plaquette with inactive variables
such that any active variables are connected.  The graph weights
$v(\Gp)$ are shown in Fig.~\ref{tJ_labels2}. Contrary to substep I,
we have to calculate the flipping probabilities of the loops according
to Eq.~(\ref{flip_tJ}), since there is no symmetry similar to the
spin-inversion symmetry in the first algorithm.
%
%
\begin{figure}[tb]
 \begin{center}
  \begin{tabular}{|c|c|c||c|c|c||c|c|c|} \hline
     \multicolumn{3}{|r||}{$\Gp$ } &
       \tabfig{psia.eps} & 
       \tabfig{psib.eps} & 
       \tabfig{psic.eps} & 
       \tabfig{psig.eps} & 
       \tabfig{psih.eps} & 
       \tabfig{psii.eps} \\\hline
     \multicolumn{3}{|r||}{$v(\Gp)$} & 
        $(1+e^{-\dt t})/2$ &
        $(e^{\dt t}-1)/2$ &
        $(1-e^{-\dt t})/2$ &
        1 & 
        1 & 
        1 \\\hline 
     \multicolumn{3}{|r||}{$\lim_{\Delta\tau\rightarrow
		 {\rm d}\tau}v(\Gp)$} & 
        1 & 
        $(t/2){\rm d}\tau$ & 
        $(t/2){\rm d}\tau$ & 
        1 & 
        1 &
        1 \\\hline\hline
%
     $\Cp$ & $w_{0}(\Cp)$ & {$w_{\scr{asymm}}(\Cp)$} & 
        \multicolumn{6}{c|}{$\Delta(\Gp,\Cp)$} \\\hline\hline
     \tabfig{phi2p.eps} \tabfig{phi2.eps} & 
     1 & 
     1 &
     1 & 0 & 1 & 0 & 0 & 0 \\
     \tabfig{phi4.eps} \tabfig{phi4b.eps}  & 
     ${\rm ch}(\dt t)$ & 
     1 &
     1 & 1 & 0 & 0 & 0 & 0 \\
     \tabfig{phi5.eps} \tabfig{phi5b.eps} & 
     ${\rm sh}(\dt t)$ & 
     1 &
     0 & 1 & 1 & 0 & 0 & 0 \\
\hline 
     \tabfig{phi3.eps} (\tabfig{phi3b.eps}) & 
     1 & 
     $e^{\Delta \tau J/2} {\rm ch}(\Delta \tau J/2)$ &
     0 & 0 & 0 & 1 & 0 & 0 \\
     \tabfig{phi4bp.eps} (\tabfig{phi4p.eps}) & 
     1 & 
     ${\rm ch}(\dt t)$ &
     0 & 0 & 0 & 1 & 0 & 0 \\
     \tabfig{phi1b.eps} (\tabfig{phi1.eps}) & 
     1 & 
     $e^{\Delta \tau J/2} {\rm sh}(\Delta \tau J/2)$ &
     0 & 0 & 0 & 0 & 1 & 0 \\
     \tabfig{phi5bp.eps} (\tabfig{phi5p.eps}) & 
     1 & 
     ${\rm sh}(\dt t)$ &
     0 & 0 & 0 & 0 & 1 & 0 \\
     \tabfig{phi2b.eps} & 
     1 & 
     1 &
     0 & 0 & 0 & 0 & 0 & 1 \\\hline
\end{tabular}
\vspace*{2ex}
\caption[*]{
        Plaquette configurations $\Cp$ and graphs $\Gp$ 
        for substep III (flip spin-up $\leftrightarrow$ hole)
        of the $t$-$J$ model.
        The solution for substep II is equivalent.
        The upper part of the figure specifies the graphs $\Gp$
        and the freezing-less solution
        $v(\Gp)$ of equation (\ref{weight_equation}). 
        The first six configurations $\Cp$ 
        and the solution $v(\Gp)$ restricted to these configurations 
        correspond to the XY-model  (free hardcore bosons).
        See also Fig.~\ref{tJ_labels1}.
}
        \label{tJ_labels2}
\end{center}
\end{figure}

In contrast to the conventional local-update worldline algorithm,
simulations can be performed in either the canonical or the grand
canonical ensemble, with either constant or variable magnetization in
the present method.  A change in the particle number or the
magnetization results from loops that wrap around the lattice in
temporal direction one or more times. If the particle number or the
magnetization should be fixed, we can simply disallow flipping these
loops without violating the detailed balance condition.  Since the
loop algorithm is no longer restricted to the subspace of a constant
spatial winding number, a negative sign may appear also for the 1D
$t$-$J$ model.  However, here the sign problem is not really a
difficulty because it becomes less significant as the system size
becomes larger.  It can also be avoided if one chooses the subspace of
constant winding number.

%
\section{Improved estimators}
\label{impr}
Let us now discuss  ``improved estimators''.\cite{improved_e} 
They  reduce the error of measured quantities
by implicitly averaging over many configurations.
In a MC simulation we construct, with the loop algorithm,
a series of $i=1,...,N$ configurations $\C^{(i)}$.
In the first step of each loop update we define a graph $\G^{(i)}$ 
which consists of a set ${\cal L}^{(i)}$ of $n^{(i)}$ loops.
From this graph we can reach any member of a set $\Gamma^{(i)}$ of
$2^{n^{(i)}}$ worldline configurations
by flipping a subset of the loops. The probability
$p(\C')$ for each of the configurations $\C' \in {\Gamma}^{(i)}$ is
determined by the loop flip probabilities $p_{\scr{flip}}$. In
the second step one configuration $\C^{(i+1)}$ will then be chosen
randomly according to these probabilities.

An improved estimator $\O_{\scr{impr}}$
for the expectation value Eq.~(\ref{observables}) 
can be constructed by averaging over the value in
each of the $2^{n^{(i)}}$ states $\C'\in{\Gamma}^{(i)}$ that can be
reached from the state $\C^{(i)}$, instead of measuring only the value
in one state $\C^{(i)}$: 
\begin{equation} \label{Eq_impr_I}
\langle \O \rangle = \langle \O_{\scr{impr}} \rangle        \,, \;\;
\O_{\scr{impr}} = \sum_{\C' \in {\Gamma}^{(i)}} \O(\C') p(\C') \,, \;\;
\ov{\O}_{\scr{impr}} ={1\over N} \sum_{i=1}^N \O_{\scr{impr}}     \,, 
\end{equation}
where the probability $p(\C')$ of the configuration $\C'$ can be
calculated as a product of the loop flip probabilities
$p_{\scr{flip}}$.  (Actually, we can choose some probability
$p'_{\scr{flip}}$ here that is different from the flip probability
used in the MC updates; it just needs to satisfy the same
detailed balance requirement as $p_{\scr{flip}}$.  Thus there is
actually a large variety of improved estimators available.)

To really gain an improvement we need to calculate the
average over $2^{n^{(i)}}$ states in a time comparable to the
time needed for a single measurement. Fortunately that is
possible.
Particularly simple improved estimators can often be found in the case
that $p_{\scr{flip}}={1\over2}$ for all loops. In that case
the above estimate simplifies to
\begin{equation}
\ov{\O}_{\scr{impr}}
= \sum_{i=1}^N 2^{-n^{(i)}}\sum_{\C' \in {\Gamma}^{(i)}} {\cal O}(\C'),
\end{equation}
as all of the states in ${\Gamma}^{(i)}$ now have the same probability
$2^{-n^{(i)}}$.

Even if the loop flip probabilities are not all equal, we can still
choose a $p'_{\scr{flip}}$ such that some loops have
$p'_{\scr{flip}}={1\over2}$, while the other loops are fixed in a
certain state.  There are many possibilities to do that. We have
chosen to fix the state of a loop with a probability of
\begin{equation}
    p_{\scr{fix}} = |2p_{\scr{flip}}-1|.
\end{equation}
If $p_{\scr{flip}}<{1\over2}$, the loop is fixed with probability
$p_{\scr{fix}}$ in the old state and if $p_{\scr{flip}}>{1\over2}$ in
the flipped state. The spins on the fixed loops are treated just as
the inactive spins.  The remaining set ${\cal F}'^{(i)}$ of
$n'^{(i)}\le n^{(i)}$ loops can then be flipped with new probabilities
$p'_{\scr{flip}}={1\over2}$.

\subsection{Simulations without sign problem}

Let us show two examples of improved estimators for the simple case of
substep I of the $t$-$J$ algorithm (or for the Heisenberg
antiferromagnet).  We provide a more detailed discussion in appendix
\ref{sec:appie}.
From Eqns.~(\ref{Eq_impr_I}) and (\ref{Cimpr}), , 
the improved estimator (multiplied by 4 for convenience)
for the spin correlation function at momentum $\pi$ is\begin{equation}
  \O_{\scr{impr}} \equiv 4 (S^z_{{\bf r},\tau} 
			    S^z_{{\bf r'},\tau'})_{\scr{impr}} = 
    \left\{ \begin{array}{lll}
             0,         & \mbox{if the spins are on different loops} \\
             1,         & \mbox{if the spins are on the same loop.} 
            \end{array}
    \right. 
\end{equation}

Remarkably, the locations of the loops thus corresponds to the
spin-spin correlation function.
The potential gain from using improved estimators is easy to see in
this case.  $\O_{\scr{impr}}$ takes only the values $0$ and $1$.  Yet
it has the same expectation value as the unimproved estimator
$\O\equiv 4 S^z_{{\bf r},\tau} S^z_{{\bf r}',\tau'} = \pm 1$.  When
$\langle \O \rangle$ is small (e.g.\ $\langle \O \rangle \sim
\exp{(-r/\xi)}$ at large $r$), then the variance of $\O$ is
\beq{var1}
  \langle \O^2 \rangle - \langle \O \rangle^2 \,=\, 
	1- \langle \O \rangle^2 \,\approx\, 1 \;,
\eeq
whereas the variance of $\O_{\scr{impr}}$ is
\beq{var2}
  \langle \O_{\scr{impr}}^2 \rangle - \langle \O_{\scr{impr}}
		 \rangle^2 \,=\,  
  \langle \O_{\scr{impr}} \rangle -  \langle \O_{\scr{impr}} \rangle^2 
  \,\approx\, \langle \O_{\scr{impr}} \rangle \equiv 
			\langle \O \rangle \ll 1 \;.
\eeq
For a given distance $r$, the gain from using the improved estimator
appears largest at small correlation length $\xi$, whereas the gain
from reducing autocorrelations with the loop algorithm is largest at
large $\xi$.  Using the improved estimator can therefore reduce the
variance, and thus the computer time required for a given accuracy, by
a large factor.  The non-improved estimator may, however, have a
sizeable amount of self-averaging from summing over all lattice sites,
which can cancel part of this gain.

An especially simple estimator can also be derived for the uniform
magnetic susceptibility
 $\langle\chi\rangle=\frac{g^2\mu_B^2\beta}{V}\left\langle
  \left({1\over{DM}}\sum_{{\bf r},\tau}S^z_{{\bf r},\tau}\right)^2
  \right\rangle$
by using 
\begin{equation}
 \sum_\tau \frac{1}{DM} \sum_{\bf r} S^z_{{\bf r},\tau}  \,=\,
  \sum_{( \scr{loops }l)}  \sum_{(({\bf r},\tau) \scr{\ in } l)} 
  \frac{1}{DM} S^z_{{\bf r},\tau}  
  \,=\, \frac{1}{2}\sum_{\scr{loops } l}  w_t(l) \;,
\end{equation}
which gives $\langle \chi \rangle = \langle \chi_{\scr{impr}} \rangle$ 
simply as the the sum of the square of the temporal winding numbers
$w_t(l)$ of the loops $l$:
\begin{equation}
\chi_{\scr{impr}} = {g^2\mu_B^2\beta\over 4V} 
	\sum_{\scr{loops }l} w_t(l)^2 \;.
\label{eq:suscm}
\end{equation}
Here $V$ is the number of spins in the lattice,
$D$ is the number of terms in the Trotter decomposition ($D=2$ 
for a nearest neighbor chain) and $M$ is the number of Trotter
time slices. Thus $VDM$ is the total number of spins in the classical
$d+1$-dimensional lattice.
In the single cluster variant, the sum over the loops in
Eq.~(\ref{eq:suscm}) is also calculated stochastically. Since there we
pick a single loop $l$ with a probability $|l|/(VDM)$ proportional to
its size $|l|$, we have to compensate for this extra factor and
obtain:
\begin{equation}
        \label{EqChiImpr1C}
        \langle \chi \rangle
        = \frac{g^2\mu_B^2 \beta}{4}
        \langle\frac{DM}{|l|} w_{t}(l)^{2}\rangle.
\end{equation}  
The improved estimators
for more general spin and charge correlations are derived in
appendix \ref{sec:appie}.

\subsection{Simulations with a sign problem}
In the case of simulations with a negative sign problem, expectation
values have to be computed according to Eq.~(\ref{Osign}).  Improved
estimators can again help here, as they reduce the variance, and thus
the error of the sign.

Let us restrict ourselves to the case of the \tJ model on a single
chain, for which
\begin{equation} \label{sign_of_config}
 \mbox{sign}(\C)=(b_{c})^{n_{x}} \left(
  (-1)^{N_{\scrscr{tot}}-1} \right)^{n_{x}} (-1)^{N_{\scrscr{neg}}}\,, 
\end{equation}
or to (possibly frustrated) spin models on any lattice, for which
\beq{signS}
 \mbox{sign}(\C) = (-1)^{N_{\scrscr{neg}}} \,.
\eeq
Here $b_{c}=+1$ for periodic and $b_{c}=-1$ for antiperiodic boundary
conditions, $N_{\scr{tot}}$ is the total particle number, $n_{x}$
denotes the number of particles hopping across the boundary and
$N_{\scr{neg}}$ are the number of plaquettes with negative weight.  
In the canonical ensemble, we can decompose the sign as 
\begin{equation} \label{sign_plaquette}
        \mbox{sign}(\C)=\prod_{\scr{plaquettes\ } p} s(\Cp),
\end{equation}
where the product extends over all plaquettes of the lattice. 
When $b_{c}(-1)^{N_{\mbox{\tiny{tot}}}-1} = -1$, 
the sign of a plaquette is defined as 
\begin{equation}
        s(\Cp)=\left\{ 
                        \begin{array}{ll}
                          -1 & \mbox{if for the \tJ model 
			             either $w(\Cp)<0$ or
                                     a particle hops across the border,
                                     but not both,} \\
                          -1 & \mbox{if for a spin model $w(\Cp)<0$}\\
                                 1 & \mbox{otherwise}
                        \end{array}
                     \right.
\end{equation}
to satisfy Eq.~(\ref{sign_plaquette}).  (When
$b_{c}(-1)^{N_{\mbox{\tiny{tot}}}-1} = 1$ it can be defined
similarly).

Improved estimators can be formulated for the sign if we can express
it as a product of signs of the loops: 
\begin{equation}
    \mbox{sign}(\C^{(i)}) 
       = s_0 \prod_{l \in {\cal L}^{(i)}} \mbox{sign}(l),
\end{equation}
where $s_0$ is the sign of the plaquettes that contain only inactive
spins.  We consider only the case where the flipping probabilities are
all $p_{\scr{flip}}=1/2$, since especially simple improved estimators
are available there:
\begin{equation}\label{ll}
    (\mbox{sign}(\C^{(i)}))_{\scr{impr}}  = s_0\cdot
  2^{-n^{(i)}}\prod_{l \in {\cal L}^{(i)}}(\mbox{sign}(l)
	+\mbox{sign}(\ov{l})).
\end{equation}
This estimator is zero if at least one loop changes its
sign when it is flipped. Again this is a very simple estimator. 

The signs of the loops are constructed in the following way: If a
plaquette contains only inactive spins, its sign contributes to
$s_0$. If only one loop threads the plaquette, the sign of the
plaquette is assigned to that loop.  The only non-trivial case is
where two loops thread a plaquette. We denote these loops by ``1'' and
``2''.  In that case the sign of the plaquette $s(\Cp)$ has to be
divided into two parts $s(\Cp)=s_{\Cp,\Gp}(1)\cdot s_{\Cp,\Gp}(2)$,
depending on the graph $\Gp$ chosen for the plaquette. This must be
done in such a way that if one or both of the loops are flipped, the
products $s_{\Cp,\Gp}(\ov{1})\cdot s_{\Cp,\Gp}(2)$,
$s_{\Cp,\Gp}(1)\cdot s_{\Cp,\Gp}(\ov{2})$ and
$s_{\Cp,\Gp}(\ov{1})\cdot s_{\Cp,\Gp}(\ov{2})$ are equal to the sign
of the plaquette with the spins on the corresponding loops
flipped. (The bars denote flipped spins on that part of the
plaquette.)  In table~\ref{sign_label} we show a solution for
$s_{\Cp,\Gp}(p)$ for substep I.  In the grand canonical ensemble we
get additional sign changes from changes of $N_{\scr{tot}}$ in
Eq.~(\ref{sign_of_config}).  They can be assigned to the loop whose
flip causes $N_{\scr{tot}}$ to change.  Overall, an assignment like
Eq.~(\ref{ll}) is possible for all three steps of the loop algorithm
for the 1D $t$-$J$ model.
For frustrated spin problems, signs appear only in single-plaquette
weights and can therefore always be assigned similar to table
\ref{sign_label}.

\begin{table}[tb] 
  \caption[*]{Assignment of signs $s_{\Cp,\Gp}(p)$ for substep I of
  the $t$-$J$ model (or for the Heisenberg model, when the bipartite
  transformation is not done).  The values of $s_{\Cp,\Gp}(p)$ for the
  plaquettes not in the table are all equal to one.
\label{sign_label}
       }
        \begin{tabular}{l|c|cccc}
        $\Cp$ & $\Gp$ & $s_{\Cp,\Gp}(1)$ & $s_{\Cp,\Gp}(\overline{1})$
                & $s_{\Cp,\Gp}(2)$ & $s_{\Cp,\Gp}(\overline{2})$ 
        \\ \hline \hline
        \tabfig{phi1.eps} \tabfig{phi1b.eps} & 
	 	\tabfig{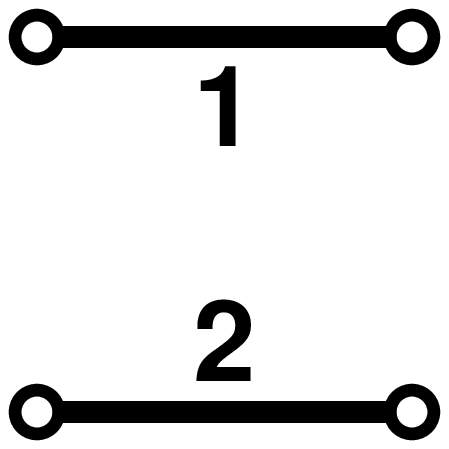} & -1 & 1 & 1 & -1 
        \\ \hline       
        \tabfig{phi3.eps} \tabfig{phi3b.eps} & 
		\tabfig{psib_no.eps} & 1 & -1 & 1 & -1 
        \\
          & \tabfig{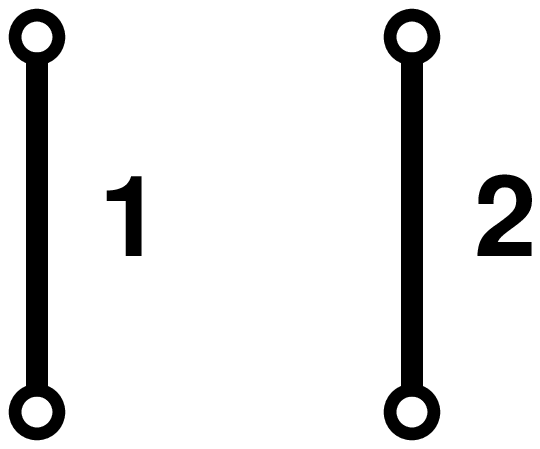} & 1 & 1 & 1 & 1 
        \\ \hline
        \tabfig{phi2.eps} \tabfig{phi2b.eps} & 
		\tabfig{psia_no.eps} & 1 & 1 & 1 & 1 
        \end{tabular}
\end{table}

If we consider different geometries for the $t$-$J$ model, such as
ladder systems or higher dimensional lattices, then $\mbox{sign}(\C)$
gets additional contributions from winding of fermion worldlines
inside the boundaries.  In this case we can still use the improved
estimators constructed above for the updates of substep I, since the
corresponding spin flips do not change the fermion winding number.
In substeps II and III more complicated improved estimators can be
constructed, at least for summing measurements over some of the
possible loop flips.  However, it is probably sufficient to have an
improved estimator for one of the substeps, in order to already obtain
most of the possible reduction in variances.

In a similar way we can also measure equal time particle-particle and
spin-spin correlations, the magnetic susceptibility and other
observables. As an example we present the improved estimator for the
uniform susceptibility in substep I of the algorithm 
(or for pure spin models).  
If {\it no loop} changes the sign of configuration $C^{(i)}$
upon flipping, then we have 
\begin{equation}
(\mbox{sign}\cdot\chi)_{\scr{impr}}^{(i)} = 
{g^2\mu_B^2\beta\over 4V}\mbox{sign}(\C^{(i)})\sum_{\scr{loops } 
  l\in{\cal L}^{(i)}} w_t(l)^2 \,.
\end{equation}
If {\it exactly two loops} $l$ and $l'$ change the sign, it is
\begin{equation}
(\mbox{sign}\cdot\chi)_{\scr{impr}}^{(i)} = 
{g^2\mu_B^2\beta\over 2V}\mbox{sign}(\C^{(i)})w_t(l)w_t(l'),
\end{equation}
and it is zero if one or more than two loops change the sign.  As this
improved estimator requires to know the sign change of a set of loops
(all those whose flip is considered in constructing the improved
estimator), a multi cluster algorithm is advantageous.  The improved
estimators for more general correlation functions are derived in
appendix \ref{sec:appie}.

\section{Results}
\label{results}
We will now discuss the performance of the new
algorithm by comparing the errors and autocorrelation times of the
local update method and the loop update with and without improved
estimators. 
We will consider four examples: a single \tJ chain, two coupled \tJ
chains, and three coupled \tJ chains (these are the first MC
simulations for coupled \tJ chains), and a frustrated Heisenberg model
on a single chain.

\subsection{Autocorrelation times}
We have determined the integrated autocorrelation times $\tio$
of our new algorithm applied to a \tJ chain.
Let us first give the details on how we have calculated
these  times.
Let $\O^{(i)}$ be the estimate of the observable $\cal O$ in the
$i$-th step of our MC procedure. It can be either the simple estimator
or the improved estimator.  As usual, we estimate the value of an
observable $\cal O$ by Eq.~(\ref{observables}) as an average over
these $N$ measurements.  (Similarly for the nominator or denominator
of Eq.~(\ref{Osign}).)
The error of the estimate is
$\sigma/\sqrt{N-1}$, where 
\begin{equation}
        \sigma^{2}=2\tio ( \overline{\O^{2}}-\overline{\O}^{2} ) \;.
\end{equation}
The autocorrelation time $\tio$
is given by the autocorrelation function $\Gamma(t)$ 
\begin{equation} \label{eq_auto_corr}
\Gamma(t)=\frac{
                \langle \O^{(t_{0}+t)} \O^{(t_{0})} \rangle
                - \langle \O^{(t_{0}+t)} \rangle \cdot
                \langle \O^{(t_{0})} \rangle
        }
        {
                \langle \O^{(t_{0})} \O^{(t_{0})} \rangle
                - \langle \O^{(t_{0})} \rangle \cdot
                \langle \O^{(t_{0})} \rangle
        }.
\end{equation}
as
\begin{equation} \label{eq_tau_int}
        \tau_{\scr{int}}^{\O} = \frac{1}{2} 
                        + \sum_{t=1}^{\infty} \Gamma(t).
\end{equation}

In the MC simulation, we have calculated $\tio$ by grouping the
$N$ measurements into $n$ bins of length $l=N/n$ and computing the bin
averages 
$\overline{\O_{b}}(l)=\frac{1}{l} 
 \sum_{j=(b-1)l+1}^{bl} \O^{(j)}$, $b=1,...,n$. 
Then we have calculated the variance of these averages
$\overline{\O_{b}}(l)$ of bin lengths $l$:
\begin{equation}
        \sigma(l)^{2}=\frac{1}{n-1}
                \sum_{b=1}^{n} \left(
                        \overline{\O_{b}}(l)-\overline{\O}
                \right)^2
\end{equation}
and the autocorrelation time can be estimated as \cite{allen_tildesley}
\begin{equation}
        \tio(l)=\frac{l \sigma (l)^{2}}{2 \sigma(1)^{2}},
\end{equation}
whose expectation value becomes equal to $\ti$ given by
Eq.~(\ref{eq_tau_int}) in the limit of $l\rightarrow \infty$.  As a
function of increasing bin length $l$, $\tio(l) $ is generally a
non-decreasing function. When statistical independence is approached,
the increase ceases and the expectation value of $\tio(l) $ approaches
a constant value.  (Note that with a finite number $N$ of
measurements, the estimate for $\tio$ will fluctuate increasingly when
$l$ is increased.)  The asymptotic constant value was our estimate for
$\tio$.  We have taken bin lengths $l=1,2,4,8,\ldots$.

For a comparison of the autocorrelation times $\tio$ we also need to
give our definitions of ``one MC step'' for both algorithms. In the
conventional plaquette flip algorithm, the lattice is subdivided into
four sublattices, which allow the simultaneous modification of all
sites of the sublattice. In a single MC step we sequentially attempted
to update all sites of a sublattice, which is generally called one
``sweep'' over the lattice. For one sweep with the loop-algorithm, we
chose one of the three steps I, II and III at random and chose graphs
$\Gp$ for all plaquettes. This results in a complete decomposition of
the lattice into loops. We then attempted to flip each loop.  All our
simulations were done in the canonical ensemble. In general we have
performed $M=2.5 \cdot 10^{6}$ MC steps for the loop algorithm and
$M=30\cdot 10^{6}$ to $70 \cdot 10^{6}$ MC steps for the plaquette
flip algorithm. Despite these very long simulation times we found
cases (only for the conventional algorithm), where $\tio(l)$ keeps
increasing as a function of the bin size $l$.  In these cases we took
$\tio(l)$ as a lower bound for Eq.~(\ref{eq_tau_int}) with the largest
$l$ where a statistically reliable estimate is still possible.

Since the value of $\tio$ depends strongly on the
observable $\O$, we have calculated $\ti$ for the internal energy, the
static charge-charge correlations
\begin{equation}
       S_{c}(k)=\frac{1}{L}\sum_{j,m}^{L}e^{ik(j-m)} 
        \langle (n_{j,\uparrow} + n_{j,\downarrow})(n_{m,\uparrow} 
        + n_{m,\downarrow}) \rangle ,
\end{equation}
and the spin-spin correlations
\begin{equation}
        S_{s}(k) =
        \frac{4}{L}\sum_{j,m}^{L}e^{ik(j-m)} 
                \langle S_{j}^{z} S_{m}^{z} \rangle.
\end{equation}

\begin{figure}[tbhp]
\begin{center}
 \psfig{figure=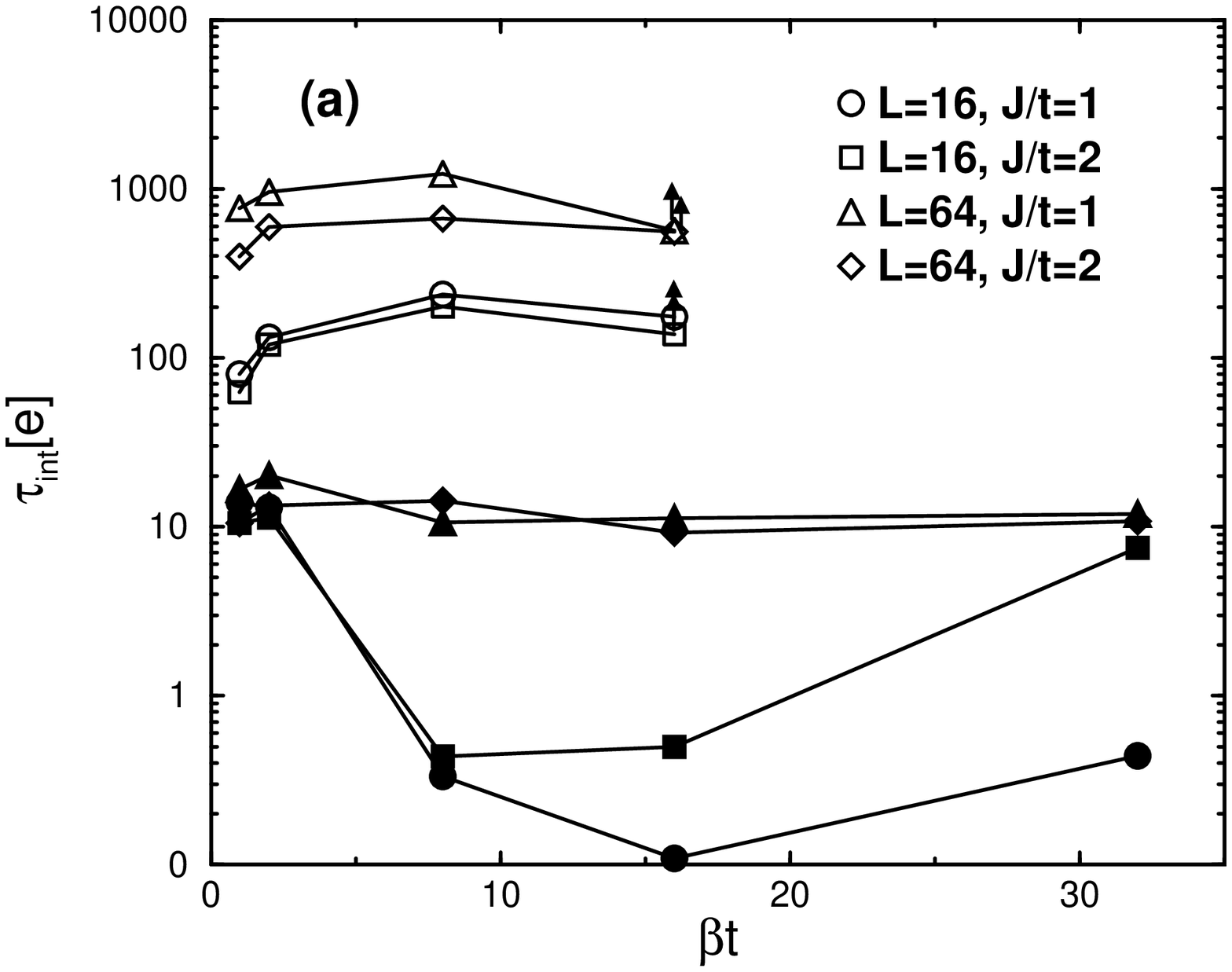,width=8cm}
 \hbox{
   \psfig{figure=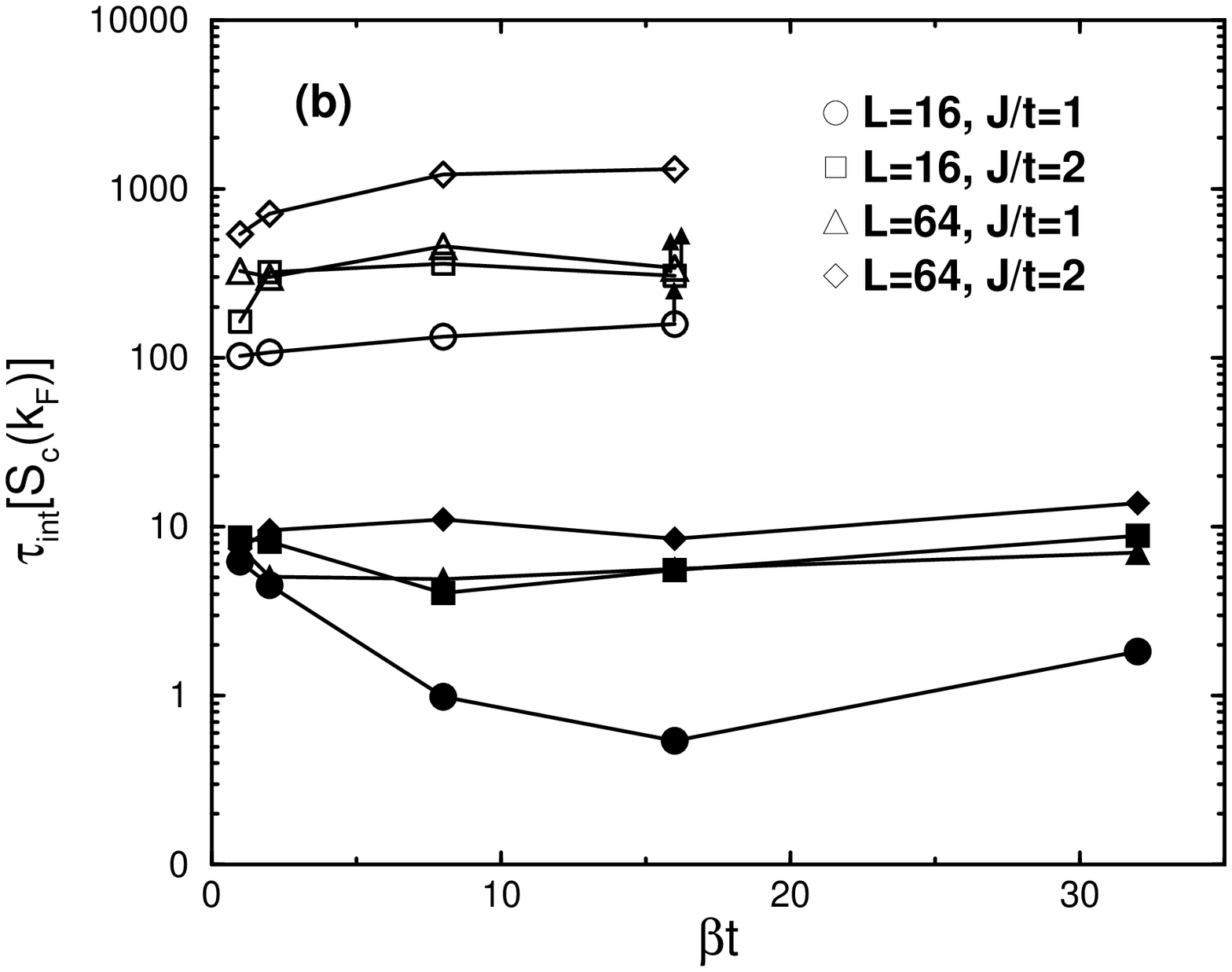,width=8cm}
   \psfig{figure=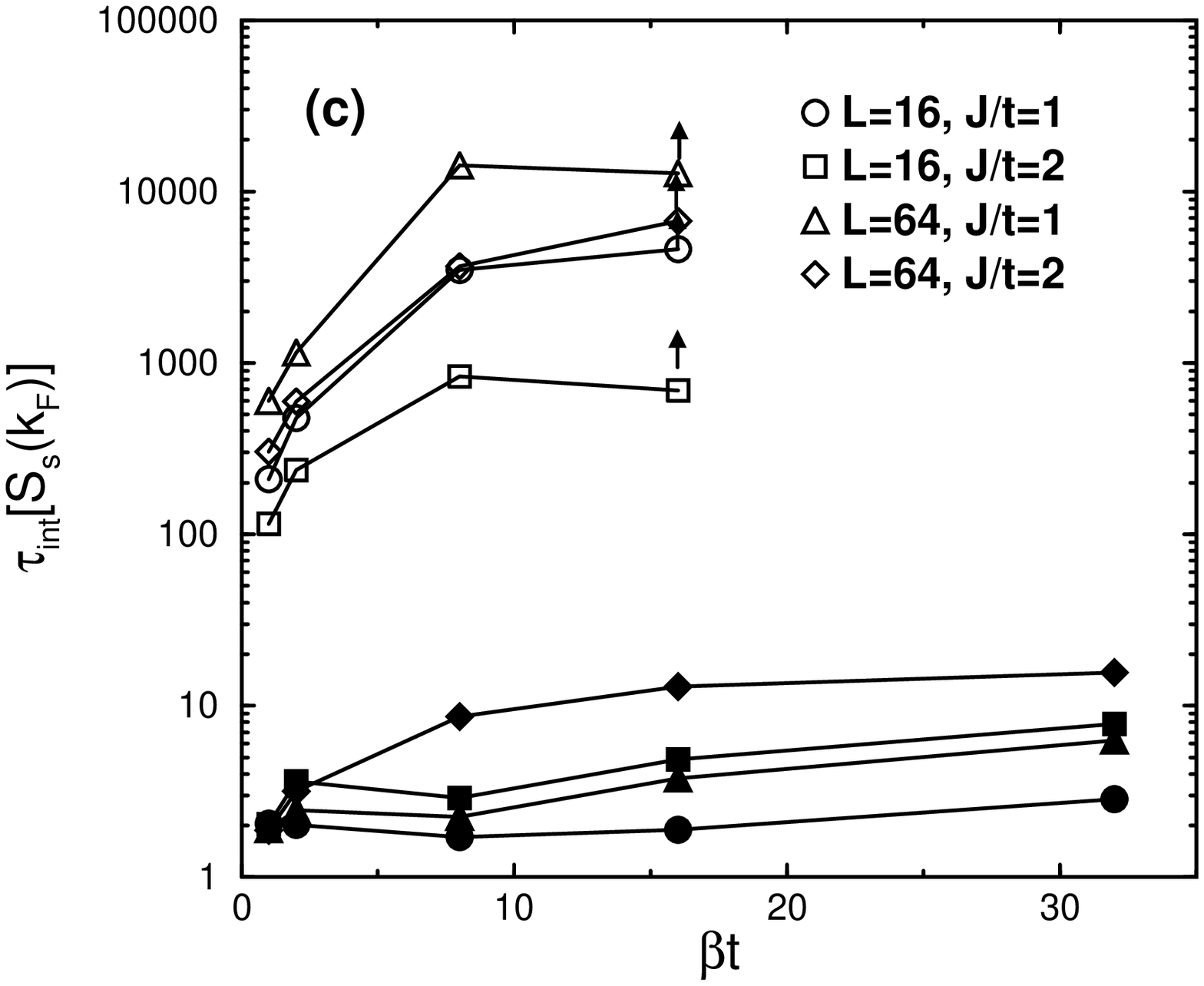,width=8cm}
      }
\end{center}
\caption{
  Integrated autocorrelation times $\tau_{\scr{int}}$ for the one
  dimensional $t$-$J$ model at quarter band filling on a lattice of
  $L=16$ sites with $\dt t=0.25$, and $L=64$ sites with $\dt t=0.125$.
  The results of the plaquette flip algorithm is shown with open
  symbols, the results of the loop algorithm with filled symbols.  For
  the loop-algorithm we took anti-periodic boundary conditions, for
  the plaquette flip algorithm we used the zero-winding boundary
  condition. Figure (a) shows $\tau_{\scr{int}}$ for the internal
  energy, (b) for charge-charge correlations $S_{c}$ at
  $k=k_{F}=\pi/4$ and (c) spin-spin correlations $S_{s}$ at
  $k=k_{F}$. The arrows denote measurements where $\tau_{\scr{int}}$
  is only a lower bound, see text for details.  The loop algorithm
  gains orders of magnitude in computational effort over the
  traditional plaquette flip algorithm, with increasing gains at low
  temperatures and for large systems.  }
\label{fig_tau}
\end{figure}

In Fig.~\ref{fig_tau} we show $\tio$ as a function of the inverse
temperature $\beta$ for a single \tJ chain.
We have performed simulations on a smaller
lattice, with $L=16$ sites and $\dt t=0.25$ and on a larger lattice
with $L=64$ sites and $\dt t=0.125$. All measurements have been
performed at quarter band filling and for ratios of $J/t=1$ and $2$.
                
For the loop algorithm we obtained values of $\ti$ between one and 15
for all observables and all parameter values.  Especially there is no
significant increase of $\ti$ with increasing $\beta$.  This is in
contrast to the conventional algorithm, where $\ti$ is between 100 and
1000 for the internal energy, but even larger than 10000 for the
spin-spin correlations.  We have to point out here that the values
of $\ti$ for the larger lattice with the conventional algorithm are
most likely  poor lower bounds of the real autocorrelation times
$\ti$, since we have not been able to reach a plateau for $\ti(l)$ in
these cases.  Obviously the loop algorithm is especially effective in
reducing the autocorrelation times for the spin-spin correlations.
Thus our new algorithm for the \tJ model works successfully,
saving orders of magnitude in computational effort.

\subsection{Improved estimators}
Next we show some results and the effect of improved estimators for
the $t$-$J$ chain, obtained with the multi-cluster algorithm. The
results of the measurements can be seen in table~\ref{Tab_tJ_Impr}.
We have considered different correlation functions, such as the
spin-spin and charge-charge correlations. For all observed quantities,
the variance is reduced with the application of improved
estimators. The variance of the improved measurements is up to
a factor of 1.7 smaller than without the use of improved measurements.
Note that in the unimproved measurements, we have measured the
correlation functions from each lattice site. Thus here we can have
large self-averaging when summing the correlation measurements over
the large lattice, which cancels part of the gain from improved
estimators.

\begin{table}[tb] 
\caption[*]{
\label{Tab_tJ_Impr}
Results for the \tJ chain: Comparison of improved (I) and unimproved
(U) measurements for the single chain $t$-$J$ model. The measured
quantities are the internal energy $e$, the charge-charge correlations
$S_{c}(k=\pi/4)$, the spin-spin correlations $S_{s}(k=\pi/4)$ and the
real-space spin-spin correlations at $r=L/8$ and $r=L/4$. We have
considered a system with periodic boundary conditions and 64 sites,
$J=t$, $\beta J=16$, $\dt t=0.125$. The number of particles $n_{p}$
are 32 and 48.  For comparison, we show in the last two rows the
results for the Heisenberg chain of the same length (hb), at $\beta
J=16$.  We performed 100000 updates for each simulation.}
\begin{tabular}{l|c|cc|cc|cc|cc|cc} 
        model & alg.
        & \multicolumn{1}{c}{$ e $} & \multicolumn{1}{c|}{error}
        & \multicolumn{1}{c}{$S_{\text{c}}(k=\frac{\pi}{4})$} 
		& \multicolumn{1}{c|}{error}
        & \multicolumn{1}{c}{$S_{\text{s}}(k=\frac{\pi}{4})$} 
		&  \multicolumn{1}{c|}{error}
        & \multicolumn{1}{c}{$\langle S_i S_{i+\frac{L}{8}} \rangle$} 
		&\multicolumn{1}{c|}{error}
        & \multicolumn{1}{c}{$\langle S_i S_{i+\frac{L}{4}} \rangle$} 
		& \multicolumn{1}{c}{error}
\\ \hline \hline
$n_{p}=32$ & U & -0.75295 & 0.000292 & 
 0.28284 & 0.00053 & 
0.78092 & 0.00202 & 
0.00533 & 0.00015 & 
0.00088 & 0.00011 \\
$n_{p}=32$ & I & -0.75254 & 0.000277 &  
0.28238 & 0.00045 & 
0.78012 & 0.00161 & 
0.00535 & 0.00012 & 
0.00086 & 0.00008 \\  \hline
$n_{p}=48$ & U & -0.81528 & 0.000322 &  
0.23708 & 0.00066 & 
0.71221 & 0.00096 & 
0.01046 & 0.00029 & 
0.00220 & 0.00021 \\
$n_{p}=48$ & I & -0.81563 & 0.000288 &  
0.23636 & 0.00056 & 
0.71135 & 0.00072 & 
0.01078 & 0.00022 & 
0.00230 & 0.00016 \\ \hline
hb & U & -1.38017 & 0.00046 &
\multicolumn{1}{c}{-} & \multicolumn{1}{c|}{-} &
0.67635 & 0.00151 &
0.04659 & 0.00064 &
0.00761 & 0.00054 \\
hb & I & -1.37965 & 0.00037 &
\multicolumn{1}{c}{-} & \multicolumn{1}{c|}{-} &
0.67480 & 0.00105 &
0.04739 & 0.00053 &
0.00797 & 0.00044 \\
\end{tabular}
\end{table}

In order to investigate the improved estimators for simulations with a
sign problem, we have simulated a simple frustrated spin system,
namely the Heisenberg chain with nearest and next-nearest neighbor
interactions. With the notations of Eq.~(\ref{Eq_Heisenberg}) the
Hamiltonian reads
\begin{equation}
 H_{JJ'}=\sum_{i}( J
 \vec{S}_{i}\cdot\vec{S}_{i+1} + J' \vec{S}_{i}\cdot\vec{S}_{i+2} ) \,.
\end{equation} 
We have implemented this model by the continuous time loop-algorithm.
For this model we have to use a finite probability
$\epsilon$ for diagonal graph segments, which implies a finite
probability for freezing (see Fig.~\ref{FigHB_labels}); otherwise the
algorithm is not ergodic (and no negative sign will appear).
The sign problem is very severe here. Even with a relatively weak
frustrating coupling of $J=10 J'$ we have obtained $\langle 
\mbox{sign}\rangle = 0.0054 \pm 0.0007$ for $\beta J=0.1$ on 20 sites.
Note that we are able to reliably measure a sign this small.  In
Tab.~\ref{Tab_JJp_ImprSign} we show the results for improved and
unimproved measurements of the sign, for different values of the
freezing factor $\epsilon$ and temperatures of $T/J=0.1$ and
$T/J=0.2$. The errors and the improvement due to the improved
estimators depends on the value of the freezing factor $\epsilon$, in
this case the optimal value is $\epsilon \approx 0.2$.  The improved
estimators perform better as the sign decreases, and the ratio of the
errors of the improved and the conventional measurements increases as
the temperature is lowered. Note that the factor of the improvement of
1.75 leads to a reduction of a factor three in CPU time.
%
%
\begin{table}[tb] 
\caption[*]{
\label{Tab_JJp_ImprSign} 
  Results for the frustrated Heisenberg chain.  Improved and
  unimproved measurements of the sign for the $J$-$J'$ Heisenberg
  model on 20 sites for $J=10 J'$ for different values of the freezing
  ratio $\epsilon$ and $\beta J=0.1$ and $\beta J =0.2$. In the last
  column, we show the ratio of the errors between the improved and the
  unimproved errors.}
        \begin{tabular}{l|c|cc|cc|c} 
	$T/J$ & $\epsilon$ & improved sign & error & 
		unimproved sign & error & ratio \\
        \hline \hline
0.1 & 0.1 & 0.00587 & 0.00109 & 0.00651 & 0.00162 & 1.48622 \\
0.1 & 0.2 & 0.00535 & 0.00075 & 0.00570 & 0.00133 & 1.76066 \\
0.1 & 0.3 & 0.00581 & 0.00084 & 0.00536 & 0.00146 & 1.74388 \\
0.1 & 0.4 & 0.00479 & 0.00084 & 0.00379 & 0.00134 & 1.60041 \\
0.1 & 0.5 & 0.00444 & 0.00105 & 0.00390 & 0.00148 & 1.41069 \\
0.1 & 0.7 & 0.00614 & 0.00223 & 0.00638 & 0.00263 & 1.17623 \\
\hline
0.2 & 0.1 & 0.08953 & 0.00201 & 0.08985 & 0.00243 & 1.20779 \\
0.2 & 0.2 & 0.08681 & 0.00151 & 0.08632 & 0.00195 & 1.29734 \\
0.2 & 0.3 & 0.08933 & 0.00152 & 0.08886 & 0.00202 & 1.32511 \\
0.2 & 0.4 & 0.08387 & 0.00166 & 0.08250 & 0.00222 & 1.33664 \\
0.2 & 0.5 & 0.08564 & 0.00219 & 0.08465 & 0.00256 & 1.16866 \\
0.2 & 0.7 & 0.08290 & 0.00372 & 0.08259 & 0.00400 & 1.07410 \\
\hline
\end{tabular}
\end{table}

\subsection{two leg \tJ ladder}
\begin{figure}[tbhp]
\begin{center}
 \hbox{
   \psfig{figure=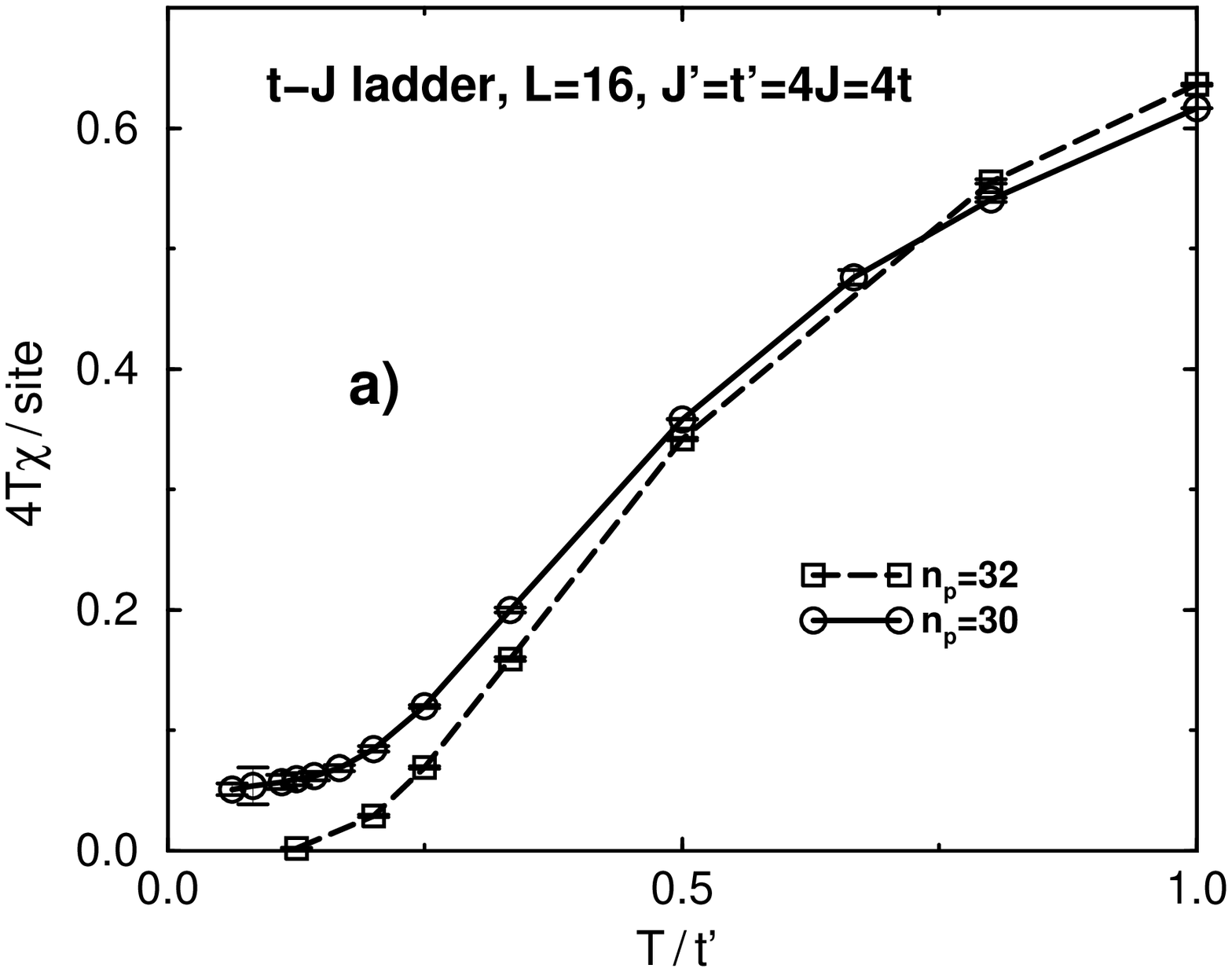,width=8cm}
   \psfig{figure=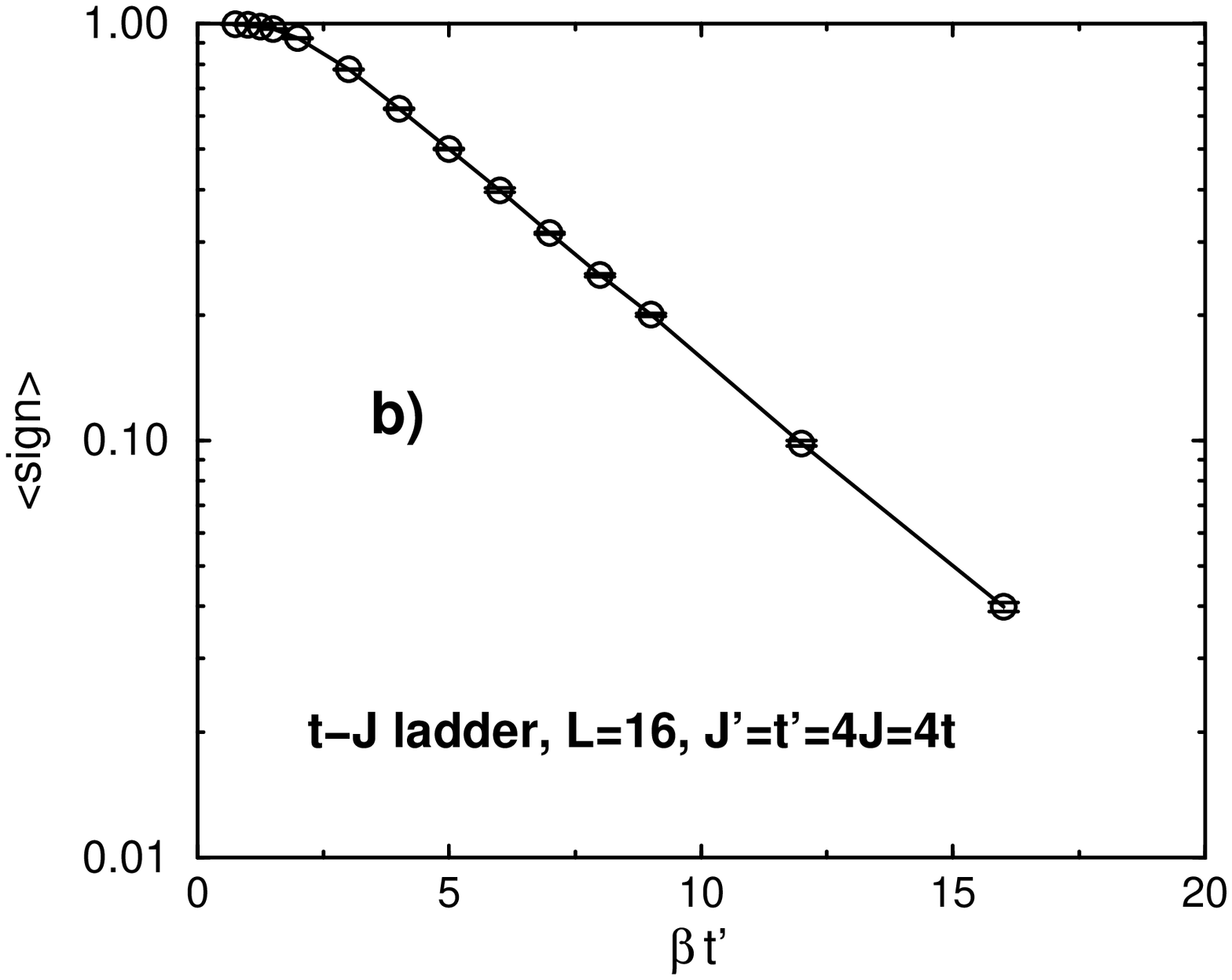,width=8cm}
      }
\end{center}
\caption{ \tJ ladder: 
    a) Magnetic Susceptibility of the $t$-$J$-ladder with $J'=t'=4J=4t$
    and two holes on 16 rungs, for temperatures down to $T=t'/16$ 
    compared to the undoped case ($n_p=32$).
    b) average sign for the doped case.
}
\label{fig_susc_lad}
\end{figure}

 As an example of the loop algorithm for the $t$-$J$
model beyond a single chain, we show the results of a calculation for
the magnetic susceptibility of the $t$-$J$ ladder.\cite{LadderReview}
In Fig.~\ref{fig_ladder}, we show a graphical representation.
\deffig{fig_ladder}{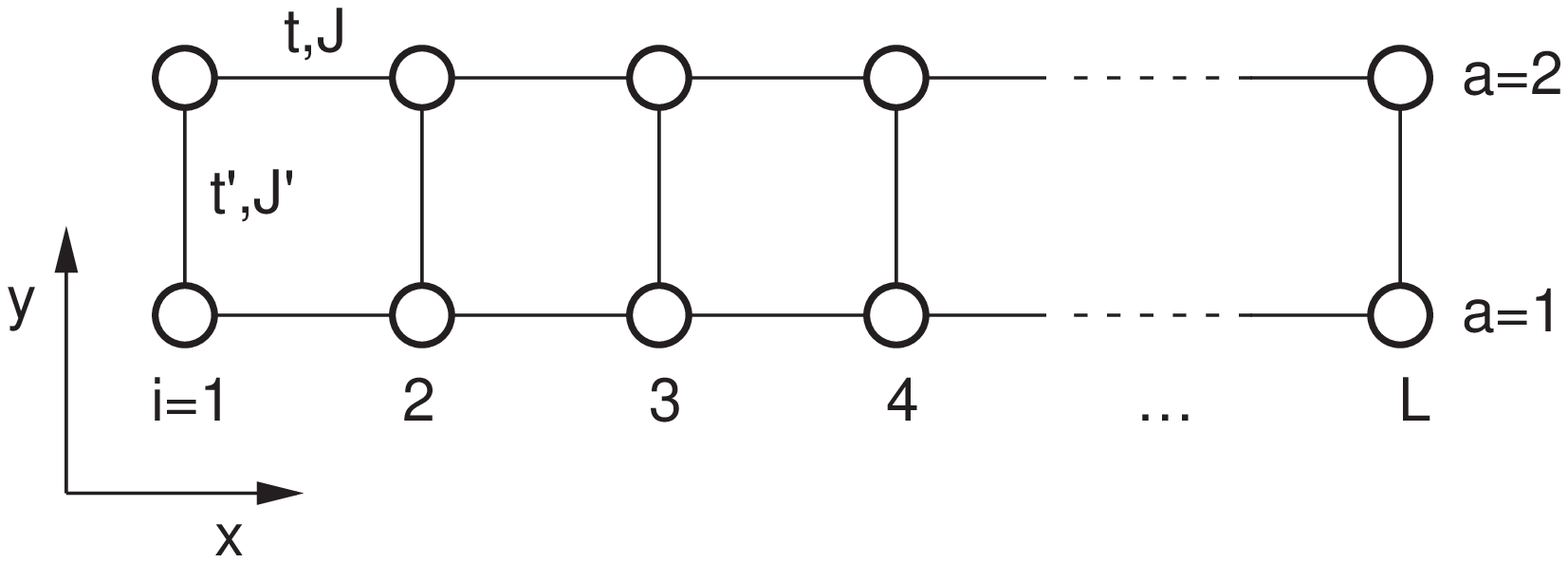}{8cm}
{Schematic picture of the $t$-$J$ ladder with two legs and $L$
rungs. The couplings along the rung are $J'$ and $t'$, those along the
ladder direction are $J$ and $t$.}  
These are the first QMC calculations for \tJ ladder systems.
For the Trotter-Suzuki breakup, we have split the Hamiltonian into
bond-terms, so that again we obtained a model on a checkerboard-like
{\em plaquette} lattice, and our loop algorithm could be applied
unchanged.
We have performed simulations  
with two holes and $J'=t'=4J=4t$, where $J'$ and $t'$ are the
interactions on each rung, and $J$ and $t$ are those along the legs. 
This parameter regime is dominated by the strong coupling limit $J'\gg
J,t$. In this limit, we have a simple and intuitive picture, following
Refs.~\onlinecite{PRBTroyer,DissTroyer}.  The undoped ladder consists
of weakly coupled singlet pairs formed on the rungs
(Fig.~\ref{fig_ladder_en}a). A single hole doped in such a ladder will
stay in either a bonding or anti-bonding orbital on a rung, while the
rest of the system will remain unchanged
(Fig.~\ref{fig_ladder_en}b). The energy of the lower lying bonding
orbital is given to first order by the cost of breaking a bond $J'$
and a kinetic energy gain of $t'$ along the rung and $t$ along the
ladder direction. Two holes on the same rung also break one bond $J'$,
but their kinetic energy gain is only of the order of $4t^{2}/J'$
(Fig.~\ref{fig_ladder_en}c).
\deffig{fig_ladder_en}{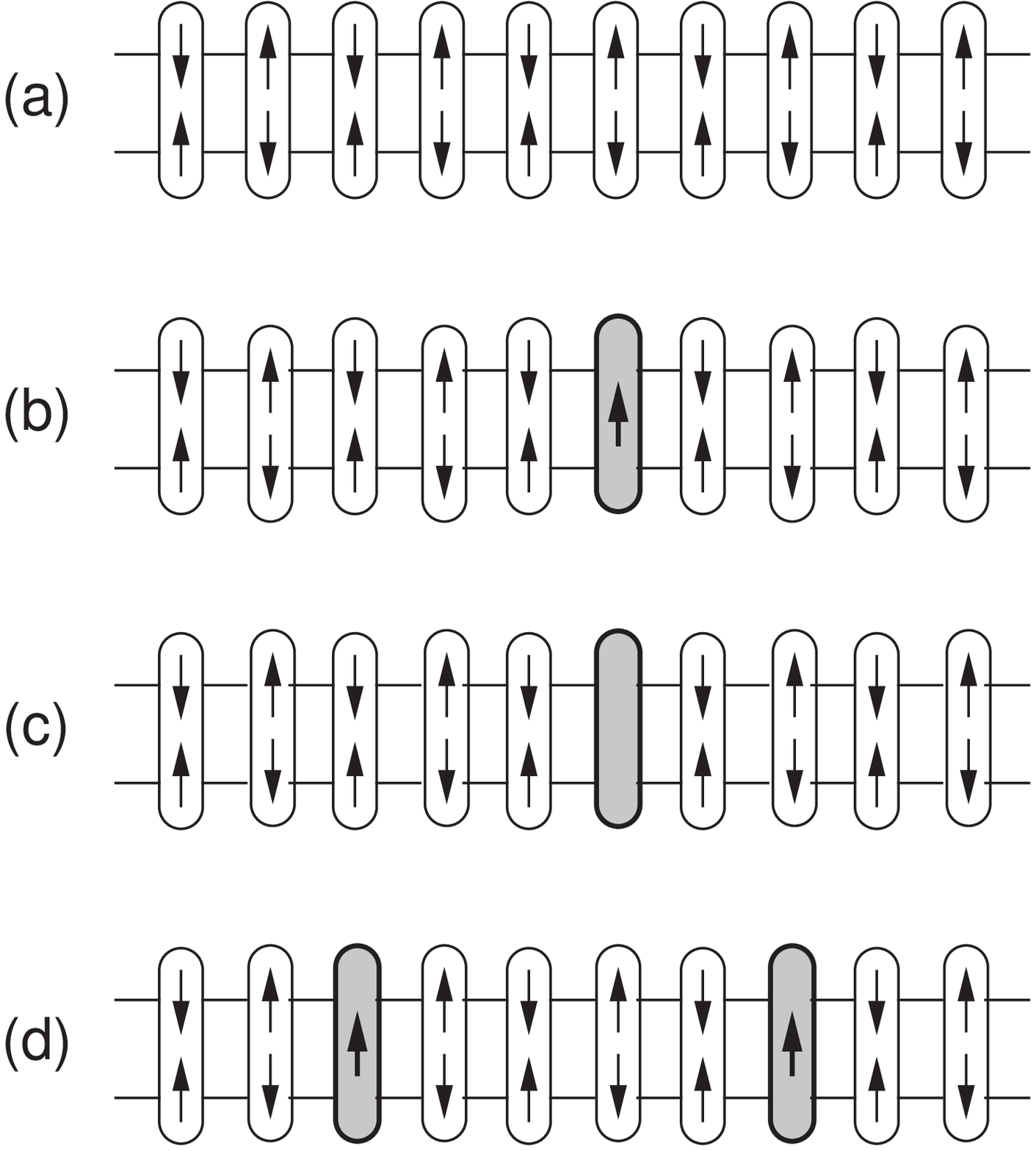}{8cm}
{Graphical representation of the low-lying states of the $t$-$J$
ladder in the strong coupling limit $J'\gg J,t$. (a) The undoped case
with a ground state energy $E(0)$. (b) A single hole goes either into
the bonding or anti-bonding orbital, the energy of the ladder with a
hole in a bonding orbital is $E(0)+J'-t'-t$ in first order. (c) Two
holes on a single rung, with $E(0)+J'$ in first order. (d) Two holes
on different rungs, with an energy $E(0)+2J'-2t'-2t$. }  
Hence the total energy of two unpaired holes in the ladder is of the
order of $E(0)+2J'-2t'-2t$ (Fig.~\ref{fig_ladder_en}d), while two
holes bound on a single rung have an energy of $E(0)+J'$ in first
order, where $E(0)$ is the energy of the corresponding Heisenberg
ladder. We can therefore expect that the two holes in the parameter
region considered in this simulation remain unpaired and thus two
of the rungs will stay in a doublet state.  The low temperature Curie
law is then given by $\chi=2\frac{s(s+1)}{3T}$. In
Fig.~\ref{fig_susc_lad} we show our results, which are in excellent
agreement with the expected limit of $4T\chi/\mbox{site}=1/16$ for two
unpaired holes and 16 rungs as $T \rightarrow 0$.  A more physically
realistic parameter region is $J'=J=t/3=t'/3$. Then two holes form a
bound state in their groundstate (Fig.~\ref{fig_ladder_en}d). 
Unfortunately the sign problem is much worse in this region.

\subsection{three leg \tJ ladder}
Several studies show that the ladders with an even number of legs
behave completely differently than those with an odd number of
legs.\cite{PRBTroyer,White,Rice} In this paragraph we will concentrate
on the 3-leg $t$-$J$-ladder. The couplings along the legs ($t$, $J$)
and the couplings perpendicular to it ($t'$, $J'$) are assumed to be
equal: $t=t'$ and $J=J'$.

At low hole doping, the three leg ladder consists of two components: a
conducting Luttinger liquid in the channel with odd parity under
reflection about the center leg, coexisting with
an insulating (i.e. undoped) spin liquid phase in the two even-parity
channels.\cite{Rice} At small doping, all holes enter the Luttinger
liquid and repel each other, while the spin liquid remains undoped. 

The energy gap $\Delta$ between the odd-parity channel and the
even-parity spin liquid states have been calculated by exact
diagonalization of very small ladders of only 3$\times$6 sites in
Ref.~\onlinecite{Rice}.  Using the QMC loop algorithm we are able to
estimate the energy gap $\Delta$ between odd and even-parity channels
for much larger ladders.  We have considered 3-leg $t$-$J$-ladders of
3$\times$40 sites doped with one hole. We have assumed periodic
boundary conditions along the ladder and set $J/t=0.5$. With this
choice of parameters, we reach temperatures down to $\beta t=7$. Below
this temperature the sign is smaller than 0.01. Note that the sign of
the MC simulations of these $t$-$J$-ladders is not sensitive to the
length of the ladder, but only to their width, the number of doped
holes and the fraction $J/t$. In the temperature range considered,
finite size effects for $L\geq 40$ are negligible.

Figure~\ref{FigThreeLeg} shows the probability $n_{\rm center\ leg}$
of the single hole to be located on the center\ leg of the 3-leg
ladder. At high temperatures the hole is uniformly distributed over
the ladder.  Therefore the density $n_{\rm center\ leg}$ is equal to
1/3. At zero temperature, however, the hole is in the lowest state of
the odd-parity channel and is dominantly on one of the outer
legs. The density $n_{\rm center\ leg}$ at $T=0$ is only $n_{\rm
center\ leg}\approx 0.2$ for a 3$\times$8 ladder.\cite{Haas}

At very low, but finite temperatures, other states with the hole in
the odd-parity channel also have non vanishing weight in the thermal
average. As these states all have odd parity, the density $n_{\rm
center\ leg}$ is suppressed and clearly smaller than $1/3$.
Fig.~\ref{FigThreeLeg} shows that at higher temperatures, $T>t/7$,
$n_{\rm center\ leg}$ is larger than $1/3$. This is caused by
admixture of higher lying even-parity channel states.  The sharp drop
of $n_{\rm center\ leg}$ below $T/t=0.5$ shows the decreasing weight
of the even-parity channel states as $T$ is lowered.  The gap $\Delta$
can be estimated from the MC data using a simple two band low-energy
model:

The two lowest lying bands of a $3 \times 8$ ladder doped with one
hole are shown in Fig. 5 of Ref.~\onlinecite{Rice}. The states
belonging to the lowest (second lowest) lying band are of odd (even)
parity. These two bands approximately have cosine forms and can be
seen as Bloch waves of two different transverse wave functions
$\phi_{\rm trans}^{\rm odd}$ and $\phi_{\rm trans}^{\rm even}$ in a
first approximation.  Then the probability of the hole to be located
on the center leg for all states in the odd (even) parity band is
constant (independent of the wave vector $k$) and equal to a value
$n_{\rm center}^{\rm odd}$ ($n_{\rm center}^{\rm even}$) which is
determined only by $\phi_{\rm trans}^{\rm odd}$ ($\phi_{\rm
trans}^{\rm even}$). From the exact diagonalization of a $3 \times 8$
ladder one sees that the approximation of a $k$ independent $n_{\rm
center}^{\rm odd}$ ($n_{\rm center}^{\rm even}$) is valid within
$10\%$, and one gets an estimate: $n_{\rm center}^{\rm odd}\approx0.2$
and $n_{\rm center}^{\rm even}\approx0.45$.\cite{Haas} Since this
situation is not supposed to change qualitatively as the length of the
ladder gets longer, the low temperature behavior of $n_{\rm center\
leg}$ can be described by this two band model also in the case of a
long ladder. Therefore, considering the density of states, the
expectation value of $n_{\rm center\ leg}$ can be calculated as a
function of the temperature $T$ and compared to the MC results. From
this one can get an estimate of the gap $\Delta$ between the odd and
even parity band in a long ladder:
\begin{equation}\label{gap}
\Delta\approx0.25(5) t = 0.5(1) J\quad (J/t=0.5).
\end{equation}
This fit is shown in Fig.~\ref{FigThreeLeg} and is reasonable at low
temperatures only. At higher temperatures other bands have to be
considered too. For this fit of the low energy model results to the MC
data all other parameters ($n_{\rm center}^{\rm odd}$, $n_{\rm
center}^{\rm even}$ and the bandwidths of the two bands) are assumed
to be equal to those of a $3 \times 8$ ladder. But even if these
paramaters are varied in a physically reasonable range, the gap
$\Delta$ hardly changes. The value obtained for $\Delta$ (Eq.~
\ref{gap}) is bigger than that of the $3\times 6$ ladder, obtained by
exact diagonalization ($\Delta=0.15t$).\cite{Rice} This difference may
result either from strong finite size effects in the small clusters or
from the fact that the low energy model described above is not so
precise in the temperature range where it was used for fitting the MC
results.

\deffig{FigThreeLeg}{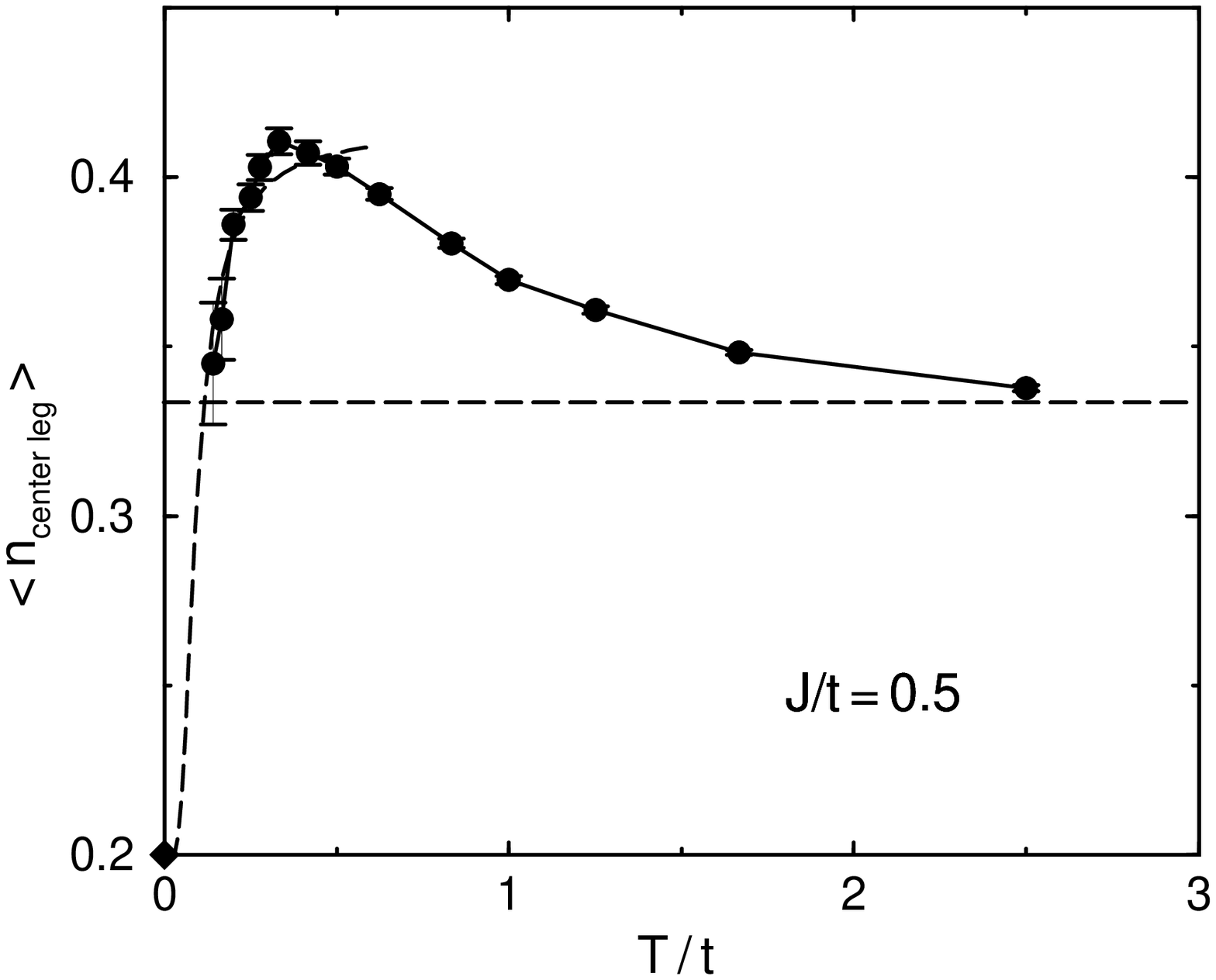}{85mm} {Temperature
  dependence of the probability of the hole to be located on the
  center leg $n_{\rm center\ leg}$ of a 3-leg $t$-$J$-ladder doped
  with one hole. The filled circles are the QMC loop data for a
  3$\times$40 cluster and the zero temperature value (diamond) is
  calculated for a 3$\times$8 cluster using exact
  diagonalization.\cite{Haas} The dashed line shows the fit calculated
  by a low-energy two band model.}

\section{Conclusions}
In this paper we have introduced a loop algorithm for simulations of
$t$-$J$ type models and discussed the use of improved estimators,
especially the use of improved estimators for models with a sign
problem.

We found many significant improvements for the loop algorithm compared
to previous local updating MC algorithms. The loop algorithm
is fully ergodic for any geometry of the lattice, without the
introduction of any additional updating procedure. 
With the loop algorithm it is possible to perform simulations in the
canonical or grand canonical ensemble, with fixed (constant winding
number) or free magnetization in a natural way.

The most important improvement of the loop algorithm is certainly the
great reduction of the autocorrelation time $\tau$. We have shown
examples in Sec.~\ref{results}, where for the parameters studied, the
reduction is up to four orders of magnitude. This gain will increase
further for larger systems and lower temperatures.  This huge
reduction of the autocorrelation times allows to study much bigger
systems at much lower temperatures than before with the same amount of
computer time.

The loop algorithm for the $t$-$J$ model can also be extended to
various other models. Different additional terms can be incorporated
easily into an overall flipping probability of the loops.  For some
new terms it might be favorable to change the weights $v(\Gp)$.  The
loop algorithm is easily adapted to other lattice geometries.  This
can be done simply by changing the underlying geometry of the lattice
in the simulation and introducing corresponding additional terms in
the Trotter decomposition.

With the loop algorithm it is also possible to perform simulations in
the continuous time limit $\dt \rightarrow 0$ (see appendix
\ref{continuous}).  Therefore, we can eliminate the errors due to the
finite time steps $\dt$ without making simulations for different
values of $\dt$ and extrapolating to $\dt=0$ afterwards. Again we can
save a large amount of computer time compared to discrete time
simulations.

The use of improved estimators further reduces the variance of
measured quantities. The introduction of improved estimators also for
models with a sign problem allows the investigation of many new
systems with this method, e.g.\ frustrated spin problems. We have
presented \tJ ladders and a frustrated Heisenberg model as examples.
The reduction of the variance by the improved estimators depends very
much on the model and the observable under consideration. For the
systems we have studied here, the improved estimators helped to reduce
the variance of the observables by about one third.

Although we can simulate much bigger systems much faster than before
with these new techniques, the sign problem still remains and limits
the application of the loop algorithm to systems where the negative
sign problem is not too severe. We have shown examples of $t$-$J$
ladder systems in Sec.~\ref{results}. Despite this drawback for higher
dimensional fermion systems, many new problems that are far beyond the
scope of previous local MC techniques can be tackled due to
the advantages of these new simulation techniques.

\acknowledgements
H.G.E. gratefully acknowledges support from BMBF (05 605 WWA 6).
B.A. was supported by an ETH-internal grant No. 9452/41-2511.5, M.T.
by the Japan Society for the Promotion of Science, and B.F. from the
Swiss Nationalfonds. Most of the calculations have been performed on
the Intel Paragon XP/S-22 MP of ETH Z\"{u}rich.

\appendix
\section{Loop algorithm in continuous imaginary time}
\label{continuous}
In this appendix we briefly review the main idea behind the continuous
time formulation of the loop algorithm \cite{beard} and how it can be
used for the $t$-$J$ model. The continuous time version depends on the
fact that the loop algorithm is well defined even in the limit
$\Delta\tau\rightarrow0$. All the probabilities for choosing graphs
$\Gp$ have well defined values in this limit.

Note that for $\Delta\tau\rightarrow 0$ the frequency of a worldline
hopping from one site to another tends to a {\it finite} limit,
because the number of time slices where such a hop can occur is
proportional to $\Delta\tau^{-1}$, and the hopping probabilities are
of the order ${\rm O}(\Delta\tau)$.
In the continuous time formulation
a configuration is therefore best specified through
the time values $\tau_i$ at which the spin
configuration changes, as well as the initial configuration at the time
$\tau=0$. This way of specifying the configuration reduces the memory
requirements by about an order of magnitude compared to a discrete
time implementation at a typical value of $\Delta\tau$.

We now describe the continuous time limit of the loop construction.
In the discrete time implementation we loop over all time slices and
decide the graph segments for each plaquette on this time slice.

In the continuous time limit we need a new procedure.  We note that
the probability for having a graph segment which forces the loop to
``jump'' to another site is proportional to the infinitesimal time
step ${\rm d}\tau$: $p=\lambda{\rm d}\tau$.  The probability for
continuing straight on is $1-{\rm O}({\rm d}\tau)$ on the other
hand. The situation is therefore equivalent to a radioactive decay
process with ``decay constant'' $\lambda$.  This decay constant
depends on the spin configuration and can change only at the time
steps $\tau_i$ where there is a change in the configuration. We have
listed the decay constants in Figs. \ref{FigHB_labels},
\ref{tJ_labels1} and \ref{tJ_labels2} together with the probabilities 
for finite $\Delta\tau$.

Instead of deciding at each infinitesimal time step ${\rm d}\tau$ if
the loop ``decays'' (i.e. jumps) to another site with probability
${\rm d}\tau$ we calculate a ``decay time'' after which the loop
``decays'' to a neighboring site. As the decay processes to the
various neighbors of a site are independent we can calculate
independent decay times for each of these ``decay channels''.
A special case are the finite number of time points where a world line
jumps to a neighbor. These are treated like in the discrete time
algorithm. There the loop has to jump to the neighboring site. This is
called a ``forced decay'' in Ref.~\onlinecite{beard}.

The loop flip probabilities also have a well-defined continuous time
limit.  In substep I they are always $1/2$, which holds even in the
continuous time case. The only nontrivial probabilities are in steps
II and III. There are two contributions from the weights
$w_{\scr{asymm}}(\Cp)$. The forced decays contribute ratios such as
\begin{equation}
w_{\scr{asymm}}(\Cp)/w_{\scr{asymm}}(\ov{\Cp}) = 
	\lim_{\Delta\tau\rightarrow 0}
{e^{\Delta \tau J/2} {\rm sh}(\Delta \tau J/2)\over {\rm sh}(\dt t)}
= {J\over 2t}
\end{equation}
or the inverse of it. For straight worldlines between two decays at
$\tau_1$ and $\tau_2$ the product over all infinitesimal weights has
to be considered.  The continuous time limit is
\begin{equation}
\lim_{M\rightarrow \infty} \prod_{j=0}^{\frac{\tau_2 - 
	\tau_1}{\beta}M} 
	w_{\scr{asymm}}(\Cp(\tau_1+j \frac{\beta}{M})).
\end{equation}
In particular
\begin{eqnarray}
\lim_{M\rightarrow \infty} \prod_{i=0}^{\frac{\tau_2 -\tau_1}{\beta}M} 
  {\rm ch}(t \frac{\beta}{M})&=&1 \\
\lim_{M\rightarrow \infty} \prod_{i=0}^{\frac{\tau_2 -\tau_1}{\beta}M} 
  e^{J\frac{\beta}{2M}}{\rm ch}(J\frac{\beta}{2M})&=&
  e^{J(\tau_2-\tau_1)/2}.
\end{eqnarray}
Thus the forced decays contribute terms like $J/2t$ or $2t/J$ and the
straight pieces just contribute the classical Ising weights of the
worldline segments.

While this continuous time algorithm is more complex to implement
than the discrete time version, it has two significant advantages.  One
advantage, mentioned already above, is that the memory requirements
are reduced by up to an order of magnitude, depending on the
implementation. This is crucial if one wants to simulate huge systems,
where memory constraints become the restricting factor.

The main advantage is however that in the continuous time algorithm
there is no systematic error associated with a finite time step
$\Delta\tau$. In the discrete time algorithm this systematic error
could be controlled by simulating for several values of the time step
$\Delta\tau$ and then extrapolating to $\Delta\tau=0$. In our
experience this need to run several simulations makes the discrete
algorithm about a factor 4-8 slower, depending on the hardware
platform and implementation.

\section{Improved estimators for correlation functions}
\label{sec:appie}
In this appendix we show improved estimators for  charge
and spin correlation functions.  First we consider the spin
correlation function 
$\langle S^z_{{\bf r},\tau} S^z_{{\bf r'},\tau'}\rangle$ 
between two spins at sites $\bf r$ and $\bf r'$ and 
at imaginary times $\tau$ and $\tau'$ respectively.  The improved
estimator is
\begin{equation}
\sum_{\C \in {\Gamma}^{(i)}} S^z_{{\bf r},\tau}(\C)
S^z_{{\bf r}',\tau'}(\C)p(\C)
\end{equation}
As each spin can be on one loop only, this sum can be simplified
substantially. If the two spins are on {\it different} loops it
is
\begin{equation}\label{B2}
\left[(1-p_{\scr{flip}})\sigma+p_{\scr{flip}}\ov{\sigma}\right]
\left[(1-p'_{\scr{flip}})\sigma'+p_{\scr{flip}}'\ov{\sigma}'\right],
\end{equation}
where $\sigma$ is the value of the $S^z_{{\bf r},\tau}$
in the original state $\C^{(i)}$, and $\ov{\sigma}$ the value in
a state where the loop containing the spin is flipped. The
flip probability of this loop is given by $p_{\scr{flip}}$.
Similarly the primed symbols refer to the other spin.
If both spins are on the {\it same} loop it is
\begin{equation}\label{B3}
\left[(1-p_{\scr{flip}})\sigma\sigma'+
p_{\scr{flip}}\ov{\sigma}\ov{\sigma}'\right]
\end{equation}
The equation for the cases where one or both spins have been fixed,
either because they are inactive in the $t$-$J$ model algorithm or
because the loop has been fixed, are straight forward.

Let us now make the above estimators more specific.  In the case of a
pure spin Hamiltonian and in substep I of the $t$-$J$ algorithm we have
$p_{\scr{flip}}={1\over 2}$ and $\ov{\sigma}=-\sigma$. In this case the
improved estimators are very simple, namely
\begin{equation}\label{Cimpr}
  (S^z_{{\bf r},\tau} S^z_{{\bf r'},\tau'})_{\scr{impr}} = 
    \left\{ \begin{array}{lll}
          0,            & \mbox{if the spins are on different loops} \\
          \sigma\sigma'\,, &\mbox{if the spins are on the same loop.} 
            \end{array}
    \right. 
\end{equation}
(Moreover, for the Heisenberg-antiferromagnet, 
we have $\sigma = +(-) \sigma' $ 
when the spins are located on the same (on different) sublattice.)
For substep II and III of the algorithm the estimators are slightly
different. There the flipping probabilities are not equal, and
$\ov{\sigma}=\pm{1\over2}-\sigma$, as we change up (down) spins into
holes and vice versa. In this case the improved estimators
Eqns.~(\ref{B2}, \ref{B3}) look more complex but can be simplified by
fixing some loops so that the remaining flipping probabilities are all
$p_{\scr{flip}}=1/2$. The spins on the fixed loops are treated just as
inactive spins.

Similar improved estimators can be used for charge-charge correlations
\begin{equation}
\sum_{\C \in {\Gamma}^{(i)}} n_{{\bf r},\tau}(\C)
n_{{\bf r}',\tau'}(\C)p(\C)
\end{equation}
with a suitable reassignment of ``$\sigma$''.  They are trivial for
substep I or pure spin models, since then only spin degrees of freedom
are changed. For steps II and III the occupation number $n$ changes to
$\ov{n}=1-n$, because these steps exchange a hole with an up or down
spin.
We see that the calculation of improved estimators of correlation
functions can be performed with similar effort as for the non-improved
estimators. 
\begin{table}[tb] 
\caption[*]{
\label{tab:iess}
Improved estimators for {\em spin correlations} in the $t$-$J$
model and in pure spin models, for simulations with a sign problem, 
from Eqns.~(\ref{ie_sign1}, \ref{ie_sign2}) in the 
case $p_{\scr{flip}}=1/2$.
        }
        \begin{tabular}{l|c|c|c}
        &only loop $l$ changes sign & both loops change sign 
                               & any other loop changes sign\\
        \hline\hline
        Step I or pure spin models&&\\
        \hline
        {\it both spins active}&&\\
         spins on different loops&0
	&$\mbox{sign}({\cal C})\sigma\sigma'$ &0 \\
         spins on same loop&0&---& 0 \\
        \hline
        {\it one or both spins inactive}&0&--- & 0\\
        \hline\hline
        Step II and III &&\\
        \hline
        {\it both spins active}&&\\
         spins on different loops
        &$\pm{1\over 4} \mbox{sign}({\cal C})(\sigma\pm{1\over4})$
        &$\mbox{sign}({\cal C})(\sigma\pm{1\over4})
	  (\sigma'\pm{1\over4})$ &0\\
        spins on same loop &
        $\mbox{sign}({\cal C})\left[\pm{1\over2}(\sigma+\sigma')
		-{1\over 8}\right]$&---&0\\
        \hline
        spin $\sigma'$ inactive & $\pm \mbox{sign}({\cal C})
		(\sigma\pm{1\over4})\sigma'$&---&0\\
        \end{tabular}
\end{table}
\begin{table}[tb] 
\caption[*]{\label{tab:iesc}
            Improved estimators for {\em charge correlations} in  
            substeps II and III of the $t$-$J$
            model algorithm, for simulations with a sign problem,
            from Eqns.~(\ref{ie_sign1}, \ref{ie_sign2}), 
            for the case $p_{\scr{flip}}=1/2$.
            (The improved estimators are trivial for substep I.)
            }
     \begin{tabular}{l|c|c|c}
        &only loop $l$ changes sign&both loops change sign 
                               & any other loop changes sign\\
        \hline
        \hline
        {\it both spins active}&&\\
        spins on different loops
        &$\mbox{sign}({\cal C}){1\over2}(n-{1\over2})$&
            $\mbox{sign}({\cal C})(n-{1\over2})(n'-{1\over2})$ &0 \\
        spins on same loop &
        $\mbox{sign}({\cal C}){1\over 2}(n+n'-1))$&--- &0 \\
        \hline
        spin $n'$ inactive & $\mbox{sign}({\cal C})(n-{1\over2})n'$
	&--- &0 \\ 
    \end{tabular}
\end{table}

For simulations with a sign problem similar improved estimators can be
derived. For the two-site spin or charge correlation functions two
different cases have to be distinguished: Both spins are on the same
loop $l$, or they are on two different loops $l$,$l'$.  If they are on
the same loop, the improved estimator is
\begin{eqnarray}\label{ie_sign1}
&&       \left[(1-p_{\scr{flip}}(l))\mbox{sign}(l)\sigma\sigma'+
p_{\scr{flip}}(l)\mbox{sign}(\ov{l})\ov{\sigma}\ov{\sigma}'\right]\\
&&\times s_0\times
\prod_{{\scr{loops}}\;i\ne l}((1-p_{\scr{flip}}(i))
	\mbox{sign}(i)+p_{\scr{flip}}(i)\mbox{sign}(\ov{i})). \nonumber
\end{eqnarray}
In the second case it is: 
\begin{eqnarray}\label{ie_sign2}
&&       \left[(1-p_{\scr{flip}}(l))\mbox{sign}(l)\sigma
	+p_{\scr{flip}}(l)
	\mbox{sign}(\ov{l})\ov{\sigma}\right] \nonumber \\
&&\times \left[(1-p_{\scr{flip}}(l'))\mbox{sign}(l')\sigma'
	+p_{\scr{flip}}(l')
	\mbox{sign}(\ov{l'})\ov{\sigma}'\right] \\
&&\times s_0\times 
\prod_{{\scr{loops}}\;i\ne l,l'}((1-p_{\scr{flip}}(i))
	\mbox{sign}(i)+p_{\scr{flip}}(i)\mbox{sign}(\ov{i})).\nonumber
\end{eqnarray}
In simulations with a sign problem it is of advantage to have all
flipping probabilities $p_{\scr{flip}}(i)=1/2$. Then the last term in
the Eqns.~(\ref{ie_sign1}) and (\ref{ie_sign2}), the product
$\prod_{i}((1-p_{\scr{flip}}(i))\mbox{sign}(i)+p_{\scr{flip}}(i)
\mbox{sign}(\ov{i}))$,
vanishes if one of the loops changes its sign, which makes the
estimators simple.  If no loops flip their sign, the improved
estimators are equivalent to the above ones
(Eqns.~(\ref{B2}, \ref{B3})).  The only other two cases with nonzero
improved estimators occur if one or both of the loops going through
the spins under consideration change their sign.  These improved
estimators are presented in tables \ref{tab:iess} and \ref{tab:iesc}.
%


\end{document}